\documentclass[journal]{IEEEtran}
\usepackage[monochrome]{color}
\usepackage{cite}
\usepackage{hyperref}
\usepackage{graphicx}
\usepackage{xcolor}

\ifCLASSINFOpdf
\else
\fi
\usepackage{amssymb,amsmath,amsthm,enumitem}
\usepackage[ruled,vlined]{algorithm2e}

\usepackage{stfloats}
\usepackage{titlesec}
\titlespacing{\section}{0pt}{1.7pt}{1.7pt}
\titlespacing{\subsection}{0pt}{1.7pt}{1.7pt}
\titlespacing{\subsubsection}{0pt}{1.7pt}{1.7pt}

\begin{document}



%
\title{TSO-DSO Operational Planning Coordination through \textcolor{blue}{``$l_1-$Proximal''} Surrogate Lagrangian Relaxation}
%
%
%

\author{Mikhail~A.~Bragin,~\IEEEmembership{Member,~IEEE,}
        and~Yury~\textcolor{blue}{V.}~Dvorkin,~\IEEEmembership{Member,~IEEE}
\thanks{M. A. Bragin is with the Department
of Electrical and Computer Engineering, University of Connecticut, Storrs,
CT, 06269, USA; e-mail: mikhail.bragin@uconn.edu.}
\thanks{Y. \textcolor{blue}{V.} Dvorkin is with the Department of Electrical and Computer Engineering at New York University’s Tandon School of Engineering, Brooklyn, NY, 11201, USA; e-mail: dvorkin@nyu.edu.}
}

\maketitle

\begin{abstract}

The proliferation of distributed energy resources (DERs), located at the Distribution System Operator (DSO) level, \textcolor{blue}{bring new opportunities as well as new challenges to the operations within the grid, specifically, when it comes to the interaction with the} Transmission System Operator (TSO). To enable interoperability, while ensuring higher flexibility and cost-efficiency, \textcolor{blue}{DSOs and the TSO} need to be efficiently coordinated.  
Difficulties behind creating such TSO-DSO coordination include the combinatorial nature of the operational planning problem involved at the transmission level as well as the nonlinearity of AC power flow within both systems. These considerations significantly increase the complexity even under the deterministic setting. In this paper, a deterministic TSO-DSO operational planning coordination problem is considered and a novel decomposition and coordination approach is developed. Within the new method, the problem is decomposed into TSO and DSO subproblems, which are efficiently coordinated by updating Lagrangian multipliers. The nonlinearities at the TSO level caused by AC power flow constraints are resolved through a dynamic linearization \textcolor{blue}{while guaranteeing feasibility through ``$l_1-$proximal'' terms}. Numerical results based on the coordination of the 118-bus TSO system with \textcolor{blue}{up to} 32 DSO 34-bus systems indicate that \textcolor{blue}{the method efficiently overcomes the computational difficulties of the problem.}

\end{abstract}


%
\IEEEpeerreviewmaketitle

\section{Introduction}
%
%
%
%
\IEEEPARstart{T}{he} widespread and rapid proliferation of distributed energy resources (DERs) such as \textcolor{blue}{Renewable Energy Resources (RES), Electric Vehicles (EVs), Storage and Demand response \cite{Papa2020, YUAN2017600, Perez2014}}, typically managed by Distribution System Operators (DSOs) \cite{Nikos}, has a far-reaching impact on the traditional power system paradigms and broad implications on the entire power grid \cite{YUAN2017600, Perez2014, Papa2020}. Specifically, the distributed generation is expected to provide services to the entire grid \cite{Mig, Mig1, Mig2, Birk}.
This proliferation is fuelled by the emerging IoT technologies enabling the integration of DERs into the grid \cite{Papa2020}, \textcolor{blue}{by decreasing technology costs \cite{Perez2014} and governmental, state, regional policies \cite{Nikos, Birk} as well as by incentives, regulatory paradigms and consumer trends \cite{Birk}}. The DSOs are expected to actively engage in TSO operations \cite{TD1, Ali}. \textcolor{blue}{Not only are DSOs expected to supply power in a cost-efficient way, but also they are expected to provide support for the TSO \cite{Grot, AlSaadi, Pilo, Alves, Puente}. Flexibility benefits benefits enabled by DERs \cite{YUAN2017600, Papa2020, Grot, EDSO1, EDSO2, Silva, Silva1, Mig, Mig1, Mig2, Florin}, which include balancing supply and demand, and voltage control \cite{YUAN2017600, EDSO1, EDSO2}, voltage fluctuation and congestion mitigation \cite{Silva}, as well as prosumer behavior of customers \cite{Papa2020}, can be achieved.} 
The cost-benefits \textcolor{blue}{have also been discussed in} \cite{Zipf, Gomez, Mig3}.  \textcolor{blue}{With high penetration of DERs, to enable efficient integration of renewable resources at TSO and DSO levels, proactive management of resources is required through appropriate TSO-DSO coordination \cite{Birk}.} 

The interoperability of TSO and DSOs becomes challenging:
1. while transmission renewable resources such as bulk wind and solar generation, DSOs' distributed energy resources (DERs) are to a large extent not observable by the TSO; 
2. the intermittent nature of renewables leads to voltage and frequency fluctuations. 
To control voltage, the consideration of AC power flow constraints is necessary within both systems. These considerations significantly increase TSO-DSO coordination problem complexity even under the deterministic  setting.


The TSO-DSO operational planning coordination is thus challenging in view of the following difficulties: 1) the presence of multiple DSOs that need to be coordinated together with the TSO; 2) the consideration of binary unit commitment (UC) decision variables at the TSO level, which lead to the drastic increase of complexity; \textcolor{blue}{3) the consideration of multiple periods (e.g., 24 hours) and inter-temporal constraints (e.g., ramp-rates); 4)} the consideration of nonlinear AC power flows at the TSO level, which contributes to the non-convexity of the problem. While the presence of AC power flow at the DSO level can be handled by using the exact second-order cone (SOC) relaxation method \cite{AC-DSO1, AC-DSO2}, the method is valid only for radial topologies and cannot be used to handle the AC power flow at the TSO level with the meshed topology of the transmission network.  At the TSO level, most research papers to date do not consider binary commitment decision variables \cite{YUAN2017600, Yuan2018, SaintPierre2017,  Polymeneas, TD4, Rossi, Cadre, Mohammadi, Papa2018, TD2, TD3, Caramanis_2015}. There have been a few attempts at coordinating TSO and several DSOs while considering binary unit commitment variables \cite{Papa2020} and \cite{TSODSOPES18, Kargarian, Nawaz}.  However, within \cite{Nikos, YUAN2017600, Rossi, Cadre, Mohammadi, Papa2018, TD2, TD3, Caramanis_2015, Papa2020, TSODSOPES18, Kargarian, Nawaz}, DC power flows were considered at the transmission level. A review of the TSO-DSO coordination models and methods can be found at \cite{Givi, Cadre}. 

\subsection{\textcolor{blue}{Structure of the Paper}}
In Section II, a TSO-DSO operational planning coordination problem is formulated while considering AC power flow within each system. 
Within TSO, generation capacity and ramp-rate, as well as AC power flow constraints (in rectangular coordinates), are considered.  Within DSOs, generation capacity constraints, second-order semi-definite relaxation for AC power flow \cite{AC-DSO1, AC-DSO2}, and nodal AC power flow balance constraints are considered.  To model the coupling between TSO and DSOs, interface power exchange constraints are considered.  


In Section III, the solution methodology is developed to coordinate the TSO and DSOs while resolving non-linearity and non-convexity difficulties introduced by AC power flow building upon the Surrogate Lagrangian Relaxation (SLR) decomposition and coordination approach \cite{SLR}, \textcolor{blue}{which naturally fits the coordination nature of the problem.} The method overcame all difficulties of previous coordination LR-based methods for solving mixed-integer programming (MIP) problems, without requiring to solve all subproblems to update multipliers and while guaranteeing smooth convergence.\footnote{The method has been further improved and has been used to successfully solve large-scale security-constrained unit commitment \textcolor{blue}{(SCUC) problems with 1000+ conventional and 80 combined cycle units} with DC power flow \cite{SALR} as well as other mixed-integer programming (MIP) problems such as assignment and scheduling \cite{SAVLR} and DNN training \cite{Deniz} while showing an advantage over such methods as branch-and-cut \cite{SALR} and ADMM \cite{SAVLR}-\cite{Deniz}.} 
The efficient coordination of TSO and DSOs within the new method rests upon the following fundamental principles: 1) exponential reduction of complexity upon decomposition, 2) contraction mapping that ensures convergence of Lagrangian multipliers and 3) dynamic linearization of AC power flow constraints with \textcolor{blue}{``$l_1-$norm'' proximal-like terms (``$l_1-$proximal'' terms, for short)} to ensure feasibility. 

The above ideas are operationalized by relaxing nodal flow balance constraints that couple nodes within TSO, and by relaxing interface power exchange constraints that couple TSO and DSOs to decompose the problem in TSO and DSO subproblems.  The resulting subproblems are solved independently of each other with much-reduced complexity and are coordinated by updating Lagrangian multipliers. Convergence is accelerated through the use of ``absolute-value" penalties (following \cite{SAVLR}) which are exactly linearized. After the linearization of penalties, the relaxed problem is still nonlinear because of the presence of nonlinear AC power flow constraints. Active and reactive power flow constraints formulated in rectangular coordinates \cite{AC_rec_1, AC_rec_2, AC_rec_3, RectCoord1, RectCoord2}  are dynamically linearized by alternatively fixing voltages at adjacent nodes while ensuring convergence and feasibility through \textcolor{blue}{``$l_1-$ proximal'' terms} which preserves the overall linearity of the relaxed problem and is amenable for the use of mixed-integer linear programming (MILP) solvers.

In Numerical Testing Section IV, a series of cases are considered to illustrate several important features of the new method. In Case Study 1, a small 4-bus system with 2 generators and 1 DSO is considered to demonstrate the convergence of multipliers. In Case Study 2, a small 9-bus system is considered \cite{matpower} to demonstrate the efficiency of linearization that enables the satisfaction of nonlinear constraints by using MILP methods.  In Case Study 3, a TSO based on topology of the IEEE 34-bus system \cite{matpower} are considered. Numerical results demonstrate that with such TSO-DSO coordination, both systems benefit and that the coordination is efficient.  
\subsection{\textcolor{blue}{Main Contributions, Novelties and the Scope of the Paper}}
\textcolor{blue}{The main contribution of the paper is the consideration of the new TSO-DSO problem formulation and the development of the novel solution methodology. Within the TSO-DSO operational planning problem formulation, the following features are considered:
\begin{itemize}
\item Binary unit commitment decisions;
\item Nonlinear AC power flow;
\item Multiple periods and inter-temporal constraints, 
\end{itemize}
\noindent although, the detailed modeling of DSO's DERs is out of the scope of the paper. 
As for the methodology, the paper is on the development of a computationally efficient ``$l_1-$ proximal'' Surrogate Lagrangian Relaxation methodology with the following distinguishing and synergistic features: 
\begin{enumerate}
    \item The drastic reduction of complexity introduced by
    \begin{enumerate}
  \item Binary unit commitment decision variables;
  \item Nonlinear AC power flow constraints;
  \item Multiple periods;
    \end{enumerate}
      through the decomposition into TSO ``zonal'' subproblems and DSO subproblems;
    \item The much reduced information exchange and privacy revelation requirements: the TSO only needs to know ``interface power exchange amounts,'' as well as the associated Lagrangian multipliers, which makes the method generalizable to address potential privacy issues; 
    \item The theoretically proved convergence of multipliers (nodal LMPs) to the optimum under realistic assumptions of the satisfaction of the simple ``surrogate optimality condition;'' 
    \item The practical implementability through the use of commercially available MIP solvers (e.g., CPLEX); at the TSO level, only an MILP solver is needed;
    \item The satisfaction of original nonlinear AC power flow constraints through the use and penalization of ``$l_1-$ proximal'' terms;
\end{enumerate}
As discussed above, points 1.a) and 1.c) have been addressed in \cite{Nikos, YUAN2017600, Rossi, Cadre, Mohammadi, Papa2018, TD2, TD3, Caramanis_2015, Papa2020, TSODSOPES18, Kargarian, Nawaz}, points 1.b) and 1.c) have been addressed in \cite{Yuan2018, SaintPierre2017,  Polymeneas, TD4, Rossi, Cadre, Mohammadi, Papa2018, TD2, TD3, Caramanis_2015} and point 1.b) has been addressed in \cite{YUAN2017600}.  Benders decomposition used within \cite{YUAN2017600} cannot be extended in a scalable way to handle 1.a) together with 1.c) because the master problem would contain all the binary variables thereby leading to high computational complexity.  In contrast, the method developed in this paper addresses 1.a)-1.c), efficiently handles complexity and is amenable for parallel/distributed processing. While per point 3), theoretical convergence has been proved in \cite{SLR, SALR, SAVLR, Deniz}, reference \cite{SLR} addressed the theoretical convergence aspects at the high level without addressing how MILP solvers can be used to solve mixed-integer nonlinear programming (MINLP) subproblems;  reference \cite{SALR} uses the method of \cite{SLR} to solve a large-scale MISO's security-constrained unit commitment (SCUC) problem with DC power flows, which is an MILP problem, and without coordination with DSOs; reference \cite{SAVLR} accelerates the convergence of the method through ``absolute-value'' penalty terms when solving MILP problems and ``$l_1-$proximal'' terms are, therefore, not used; reference \cite{Deniz} extends the method to solve nonlinear problems within a difference problem context of deep neural network training. Points 4) and 5) allow the use of MILP solvers without sacrificing the overall feasibility of the MINLP Unit Commitment (UC) problem under consideration.}

\textcolor{blue}{The game-theoretical aspects, $N-1$ criteria, privacy, cybersecurity, parallel processing/distributed coordination, and stochastic optimization considerations are beyond the scope of the paper. Although, the framework of this paper will provide the necessary plug-and-play, theoretical and computational capabilities as well as lays out the communication foundations for future developments to address the above considerations. Moreover, from the ISO's perspective, the consideration of security-constrained unit commitment (SCUC) is important. To this end,  SCUC with DC power flow and 1000+ units \cite{SALR} has been efficiently solved by a previous version of the SLR method. With the capabilities of the new method developed in this paper to handle nonlinear AC power flow constraints, such problems as SCUC + AC OPF can also be solved efficiently. 
}
\section{TSO-DSO Optimization Model}

In subsection A, the TSO problem formulation is presented. In subsection B, the DSO problem formulations is presented.  In subsection C, the overall TSO-DSO coordination problem is developed.  

\subsection{TSO Model} \label{sec:tso}

Consider a transmission network with meshed topology operated by a transmission system operator (TSO) with a set of interconnections with several distribution networks (DSOs). 
Let  $\mathcal{T}$ be the lookahead horizon: $\mathcal{T} = \{1,2, ... , T\}.$ \textcolor{blue}{The planning horizon is typically $T=24$ and time resolution is $1$ hour.} Within the TSO, let $\mathcal{B}^{T}$ be a set of buses indexed by $b^T$, $\mathcal{I}^{T}_{b^T}$ be a set of generators at bus ${b^T}$ indexed by $i$, $\mathcal{J}^{T}$ ($\mathcal{J}^{T} \subset \mathcal{B}^{T}$) be a set of interconnections/root buses, whereby distribution networks are connected to the transmission network, indexed by $j$, $\mathcal{L}^{T}$ be a sets of transmission lines indexed by $l$.  

\noindent \textbf{Objective.} The TSO aims to minimize the total  generation cost as well as the total cost of power exchange at the interface of the TSO and DSOs: 
\vspace{-1mm}
\begin{align}
& \min_{\substack {\textcolor{blue}{\mathbf{F}^T}, \mathbf{G}^T,\mathbf{P}^T, \\ \mathbf{Q}^T,  \textcolor{blue}{\mathbf{V}^T}, \textcolor{blue}{\mathbf{X}^T}}} \Big\{O^T(\mathbf{G}^T,\mathbf{P}^T,\mathbf{Q}^T)\Big\}  = \label{tsoobjective} 
\end{align}
\vspace{-3mm}
\begin{align}
& \hspace{0.1cm} \nonumber \min_{\substack {\textcolor{blue}{\mathbf{F}^T}, \mathbf{G}^T,\mathbf{P}^T, \\ \mathbf{Q}^T,  \textcolor{blue}{\mathbf{V}^T}, \textcolor{blue}{\mathbf{X}^T}}}  
\begin{Bmatrix}
\displaystyle\sum_{t \in \mathcal{T}, b^T \in \mathcal{B}^{T}, i \in \mathcal{I}^{T}_{b^T}} \!\!\Bigg( C_{i,t}^{T,p}  g^{T,p}_{i,t} \!+\! C_{i,t}^{T,q}  g^{T,q}_{i,t} \Bigg)  \\
\:\: + \displaystyle\sum_{t \in \mathcal{T},j \in \mathcal{J}^{T}} \Bigg( C_{j,t}^{p} p^{T}_{j,t} + C_{j,t}^{q} q^{T}_{j,t}\Bigg)    
\end{Bmatrix}
\!,
\end{align}


\noindent where $\mathbf{G}^T = \big\{g_{i,t}^{T,p}, g_{i,t}^{T,q}\big\}$ is a vector of active and reactive\footnote{\textcolor{blue}{Reactive power has been used to control the TSO voltage \cite{Marten, Nik} as well as for market competitiveness \cite{Gil} and completeness \cite{Lipka}.}} generation levels with the corresponding generation costs $C_{i,t}^{T,p}$ and $C_{i,t}^{T,q}$,  $\mathbf{P}^T = \big\{p^{T}_{j,t}\big\}$ is a vector of active interface power exchange amounts with the corresponding DSO $j^{th}$ bids $C_{j,t}^{p}$, and $\mathbf{Q}^T = \big\{q^{T}_{j,t}\big\}$ is a vector of reactive interface power exchange amount with the corresponding DSO $j^{th}$ bids $C_{j,t}^{q}$.  \textcolor{blue}{TSO-DSO interface power exchange are assumed to be bi-directional with the corresponding amounts} of $p^{T}_{j,t}$ and $q^{T}_{j,t}$, which are positive if power exchange flow is from DSO to TSO, i.e., TSO buys power, and negative, otherwise.     

\textcolor{blue}{Other decision variables include $\mathbf{X}^T = \big\{x_{i,t}\big\}$ - a vector of binary unit commitment decision variables, $\mathbf{F}^T = \big\{f_{l,t}^{T,p}, f_{l,t}^{T,q}\big\}$ - a vector of active ($p$) and reactive ($q$) power flows and $\mathbf{V}^T = \big\{v_{b^T,t}^{T,Re}, v_{b^T,t}^{T,Im}\big\}$ - a vector of real ($Re$) and imaginary ($Im$) voltages.}

The optimization \eqref{tsoobjective} is subject to the following constraints:

\noindent \textbf{Generation Capacity Constraints.} Active $g^{T,p}_{i,t}$ and reactive $g^{T,q}_{i,t}$ generation levels of unit $i$ are constrained as follows:
\vspace{-2mm}
\begin{flalign}
& \underline{G}^{T,p}_i \cdot x_{i,t} \leq   g^{T,p}_{i,t} \leq \overline{G}^{T,p}_i \cdot x_{i,t}, \label{tso_eq2}   
\end{flalign}
\vspace{-6mm}
\begin{flalign}& \underline{G}^{T,q}_i \cdot x_{i,t} \leq   g^{T,q}_{i,t} \leq \overline{G}^{T,q}_i \cdot x_{i,t}, \label{tso_eq2b}   
\end{flalign}

\noindent 
where the minimum and maximum active power limits are $\underline{G}^{T,p}_{i}$ and $\overline{G}^{T,p}_{i}$, while the minimum and maximum reactive power limits are $\underline{G}^{T,q}_{i}$ and $\overline{G}^{T,q}_{i}$.

\noindent \textbf{Ramp-Rate Constraints.} Ramp-rate constraints require that the change of power generation levels between two consecutive time periods does not exceed ramp rates $R_i^{p/q}$: 
\begin{flalign}
& \!\!\!g_{i,t}^{T,p} \! -\! g_{i,t-1}^{T,p} \! \leq \! R_i^p  \!\cdot \!  x_{i,t-1} \! + \! \left(\underline{G}_{i}^{T,p} \!+\!\frac{R_i^p}{2}  \right) \!\cdot\!  \left(x_{i,t}\!  -  \!x_{i,t-1} \right),  \label{tso_eq3} 
\end{flalign}
\vspace{-3mm}
\begin{flalign}
& \!\!\! g_{i,t}^{T,q} \! -\!g_{i,t-1}^{T,q} \! \leq \! R_i^q  \!\cdot\!   x_{i,t-1} \! + \! \left(\underline{G}_{i}^{T,q} \!+\!\frac{R_i^q}{2}  \right) \!\cdot \!\left(x_{i,t} \! - \! x_{i,t-1} \right),  \label{tso_eq3a} 
\end{flalign}
\vspace{-3mm}
\begin{flalign}
& \!\!\! g_{i,t-1}^{T,p}\!-\!g_{i,t}^{T,p}\! \leq \!R_i^p \!\cdot \!x_{i,t} +  \left(\underline{G}_{i}^{T,p}\!+\!\frac{R_i^p}{2} \right) \!\cdot\! \left(x_{i,t-1} - x_{i,t} \right), \label{tso_eq4} 
\end{flalign}
\vspace{-3mm}
\begin{flalign}
&  \!\!\!g_{i,t-1}^{T,q}\!-\!g_{i,t}^{T,q} \!\leq \!R_i^q \!\cdot \!x_{i,t} +  \left(\underline{G}_{i}^{T,q}\!+\!\frac{R_i^q}{2} \right) \!\cdot\! \left(x_{i,t-1} - x_{i,t} \right). \label{tso_eq4a} 
\end{flalign}

\noindent For a complete set of tightened ramp-rate constraints, refer to \cite{BingRamp}.


\noindent \textbf{Nodal Power Flow Balance Constraints.} For every node $b^T$, the net active/reactive power generated and transmitted to the node should be equal to the net power consumed and transmitted from node $b^T$: 
\begin{flalign}
& \sum_{i \in \mathcal{I}^{T}_{b^T}} g_{i,t}^{T,p}+\!\sum_{\substack{l=1: \\r(l)=b^T}}^L f_{l,t}^{T,p} + p^{T}_{b^T,t} = L_{b^T,t}^{T,p} + 
 \! \sum_{\substack {l=1: \\ s(l)=b^T}}^L \! f_{l,t}^{T,p}.  \label{tso_eq7} 
\end{flalign}

\noindent  If bus $b^T$ does not contain generators, then $\sum_{i \in \mathcal{I}^{T}_{b^T}} g_{i,t}^{T,p} = 0$ if bus $b^T$ is not a root bus ($b^T \notin \mathcal{J}^{T}$), then interface active power exchange levels are zero $p^{T}_{b^T,t} = 0$ and if bus $b^T$ does not contain load, then $L_{b^T,t}^{T,p}$ = 0.  The nodal power flow balance constraints for reactive power are similarly defined:  
\begin{flalign}
& \sum_{i \in \mathcal{I}^{T}_{b^T}} g_{i,t}^{T,q}+\!\sum_{\substack{l=1: \\r(l)=b^T}}^L \!\!f_{l,t}^{T,q} + q^{T}_{b^T,t} = L_{b^T,t}^{T,q} + 
\!  \sum_{\substack {l=1: \\ s(l)=b^T}}^L \!\!f_{l,t}^{T,q}.  \label{tso_eq7a} 
\end{flalign}
\noindent \textbf{AC Power Flow Constraints in Rectangular Coordinates.} Following \cite{AC_rec_1, AC_rec_2, AC_rec_3, RectCoord1, RectCoord2}, AC power flow is modeled in rectangular coordinates by using complex voltages $v_{b^T,t}^T \! = \!v_{b^T,t}^{T,Re}\! +\! j \cdot v_{b^T,t}^{T,Im}$.  

In the complex plane, complex voltages can be represented as row vectors $v_{b^T,t}^T = \left(v_{b^T,t}^{T,Re}, v_{b^T,t}^{T,Im}\right)$ and power flows can be written as:\footnote{Here superscript ``T" indicates voltage at the TSO level, while ``T" outside of the parentheses indicates the ``transpose."}  
\begin{flalign}
& f_{l,t}^{T,p}=v_{s(l),t}^T \!\cdot\! \begin{pmatrix} g_{s(l),r(l)}\,\,\,\,    \mbox{-} b_{s(l),r(l)} \\
b_{s(l),r(l)}\,\,\,\,\,\,    g_{s(l),r(l)}  \end{pmatrix} \!\cdot \!\left( v_{r(l),t}^T \right)^T, \label{tso_eq10} 
\end{flalign}
\vspace{-3mm}
\begin{flalign}
& f_{l,t}^{T,q}=v_{s(l),t}^T \!\cdot\! \begin{pmatrix} \mbox{-}b_{s(l),r(l)}\,\,\,   \mbox{-}g_{s(l),r(l)} \\
\,\,\, g_{s(l),r(l)}\,\,\,   \mbox{-}b_{s(l),r(l)} \end{pmatrix} \!\cdot\! \left( v_{r(l),t}^T \right)^T . \label{tso_eq11} 
\end{flalign}

\noindent Here $b_{s(l),r(l)}$ is susceptance and $g_{s(l),r(l)}$ is conductance of line $({s(l),r(l)})$.  Node $s(l)$ denotes the ``sending" node of line $l$, and $r(l)$ denotes the ``receiving" node of the line $l$.  

\noindent \textbf{Voltage Restrictions.} The complex voltage within each node $b^T$ is subject to the following restrictions:
\begin{flalign}
& \underline{v_{b^T}^T} \leq \sqrt{\big(v_{b^T,t}^T\big) \cdot  \big(v_{b^T,t}^T\big)^T}, \label{tso_eq12a}
\end{flalign}
\vspace{-5mm}
\begin{flalign}
& \sqrt{\big(v_{b^T,t}^T\big) \cdot \big(v_{b^T,t}^T\big)^T} \leq \overline{v_{b^T}^T}. \label{tso_eq12b}
\end{flalign}
\noindent \textbf{Transmission Capacity Constraints.} Power flows in each line $l$ satisfy the following transmission capacity constraints:
\begin{flalign}
& \sqrt{\big(f_{l,t}^{T,p}\big)^2 + \big(f_{l,t}^{T,q}\big)^2}\leq \overline{f}_{l}. \label{tso_eq13}
\end{flalign}
\subsection{DSO Model}\label{sec:dso}
Consider a distribution network with a radial topology\footnote{\textcolor{blue}{Radial topology is considered for simplicity. Power flows within DSOs with meshed topologies can be modeled in the same way as within the TSO as described above.}} operated by a distribution system operator (DSO) and connected to a TSO's root bus $j$.  Within each DSO, let $\mathcal{B}^{D}$ be a set of buses indexed by $b^D$, $\mathcal{I}^{D}_{b^D}$ be a set of generators at bus $b^D$ indexed by $i$, $\mathcal{L}^{D}$ be a set of transmission lines indexed by $l$.

\noindent \textbf{Objective.} The DSO aims to minimize the total generation cost and to maximize the profit from selling power to the TSO. Therefore, the following sum of the two objectives is minimized as:
\begin{align}
& \min_{\textcolor{blue}{a^{D}_j}, \textcolor{blue}{f^{D}_j},g^{D}_j,p^{D}_j,q^{D}_j,\textcolor{blue}{v^{D}_j}} \Big\{O^D_j \left(g^{D}_j,p^{D}_j,q^{D}_j; \mathbf{\Lambda}_j \right)\Big\} = \label{dso_obj} 
\end{align}
\vspace{-3mm}
\begin{align}
& \hspace{0.1cm} \nonumber \min_{\substack {\textcolor{blue}{a^{D}_j}, \textcolor{blue}{f^{D}_j},g^{D}_j, \\ p^{D}_j,q^{D}_j,\textcolor{blue}{v^{D}_j}}}  
\begin{Bmatrix}
\displaystyle\sum_{i \in \mathcal{I}^{D}_{b^D}\!,  t \in \mathcal{T}}\!\!
 \Bigg(\!C_{i,j,t}^{D,p} \!\cdot \!g_{i,j,t}^{D,p} \!+\!  C_{i,j,t}^{D,q} \!\cdot\! g_{i,j,t}^{D,q} \!\Bigg) \!\!\\
-\displaystyle\sum_{t \in \mathcal{T}} \Bigg(\lambda^p_{j,t} \cdot p^{D}_{j,t} \!+\!  \lambda^q_{j,t} \cdot q^{D}_{j,t}   \Bigg)  
\end{Bmatrix}
\!, j \!\in  \!\mathcal{J}^T, 
\end{align}

\noindent where $\mathbf{\Lambda}_j = \big\{\lambda_{j,t}^{p}, \lambda_{j,t}^{q}\big\}$ denote the locational marginal prices at the transmission bus $j$, which are  a set of Lagrangian multipliers corresponding to nodal power flow balance constraints \eqref{tso_eq7} and \eqref{tso_eq7a}.  Multipliers are the wholesale LMPs at the transmission root bus $j \in \mathcal{J}^{T}$ with an interconnected distribution system $j$. Within the distribution network $j$, each conventional generator $i$ at hour $t$ produces active power denoted by $g_{i,j,t}^{D,p}$ and reactive power denoted by $g_{i,j,t}^{D,q}$.  The corresponding active and reactive power generations costs are $C_{i,j,t}^{D,p}$ and $C_{i,j,t}^{D,q}$. Accordingly, the first summation within \eqref{dso_obj} accounts for the active and reactive power generation cost of conventional generators located within the distribution system $j$. The second summation within \eqref{dso_obj} accounts for the cost of transactions performed by the DSO in the wholesale electricity market in response to price signals $\lambda^p_{j,t}$ and $\lambda^q_{j,t}$. 
\textcolor{blue}{Similar to the TSO's case, interface power exchange amounts} are $p^{D}_{j,t}$ and $q^{D}_{j,t}$, which are positive if power exchange flow is from DSO to TSO, i.e., DSO sells power, and negative, otherwise.

\noindent \textbf{Generation Capacity Constraints.} Active $g^{D,p}_{i,j,t}$ and reactive $g^{D,q}_{i,j,t}$ generation levels of unit $i$ are constrained as:
\begin{align}
&  \underline{G}^{D,p}_{i,j} \leq g^{D,p}_{i,j,t} \leq \overline{G}^{D,p}_{i,j},  \;\; \underline{G}^{D,q}_{i,j} \leq  g^{D,q}_{i,j,t} \leq \overline{G}^{D,q}_{i,j},\label{dso_eq2}
\end{align}
where the minimum and maximum active power limits are $\underline{G}^{D,p}_{i,j}$ and $\overline{G}^{D,p}_{i,j}$, while the minimum and maximum reactive power limits are $\underline{G}^{D,q}_{i,j}$ and $\overline{G}^{D,q}_{i,j}$.  

\noindent \textbf{Second-Order Cone Relaxation of AC Power Flow.} Since distribution systems are assumed to have a radial topology, AC power flows are modeled by using an exact second-order cone (SOC) relaxation following  \cite{AC-DSO1, AC-DSO2}.  The following constraint captures relationships among active $\big(f^{D,p}_{l,j,t}\big)^2$ and reactive $\big(f^{D,q}_{l,j,t}\big)^2$ power flows squared as well as current squared $a_{l,j,t}$ in line $l$, and voltage squared $v_{s(l),j,t}$ at the sending end $s(l)$ of line $l$ as:
\begin{align}
& \big(f^{D,p}_{l,j,t}\big)^2 + \big(f^{D,q}_{l,j,t} \big)^2 \leq  \hspace{0.0cm} v_{s(l),j,t} \cdot a_{l,j,t}.  \label{dso_eq3} 
\end{align}
 
 \noindent To capture voltage drops across line $l$ between voltages squared at sending $s(l)$ and receiving $r(l)$ buses, the following constraint is used:
\begin{flalign}
& v_{r(l),j,t} - 
v_{s(l),j,t} = \\  \nonumber
& 2 \cdot \left(R_{l,j} \cdot f^{D,p}_{l,j,t} + X_{l,j} \cdot f^{D,q}_{l,j,t} \right) - 
a_{l,j,t} \cdot \left(R_{l,j}^2 + X_{l,j}^2 \right), 
\end{flalign}
\noindent where parameters $R_{l,j}$ and $X_{l,j}$ are the reactanace and impedance of line $l$. 

 Since power flow at sending and receiving buses of each line $l$ differs due to losses incurred by transmission, the apparent power flow limit $\overline{S}_{l,j}$  is enforced for the sending and receiving buses separately within the following two sets of constraints:   
\begin{flalign}
& \big(f^{p}_{l,j,t} \big)^2 + \big(f^{q}_{l,j,t} \big)^2 \leq \big(\overline{S}_{l,j}\big)^2,    \label{dso_eq5} 
\end{flalign}
\begin{flalign}
& \big(f^{D,p}_{l,j,t}  - a_{l,j,t} R_{l,j} \big)^2 \!+\!  \big(f^{D,q}_{l,j,t}\! -\! a_{l,j,t} X_{l,j} \big)^2  \label{dso_eq6}  \leq \big(\overline{S}_{l,j}\big)^2.  
\end{flalign}

\noindent \textbf{Voltage Restriction Constraints.} The bus voltages squared are limited by $\underline{v^D_{b^D,j}}$ and $\overline{v^D_{b^D,j}}$ as:
\begin{flalign}
& \underline{v^D_{b^D,j}}  \leq v_{b^D,j,t} \leq \overline{v^D_{b^D,j}}. \label{dso_eq7} 
\end{flalign}

\noindent \textbf{Nodal Power Flow Balance Constraints.} With the newly added load $L_{b^D,j,t}^{D,p}$, generation levels $g_{i,j,t}^{D,p}$ and interface power exchange amount $p^{D}_{j,t}$, the nodal power balance is enforced following \cite[eq. (3)]{AC-DSO2} as:
\begin{flalign}
&  \sum_{l|s(l)=b^D} f_{l,j,t}^{D,p} -   \left(f_{l,j,t}^{D,p} - a_{l,j,t} \cdot R_{l,j} \right)_{l|r(l)=b^D} -\label{dso_eq8} \\ & \hspace{0.2cm} \sum_{i \in \mathcal{I}^{D}_{b^D}} g_{i,j,t}^{D,p} + L_{b^D,j,t}^{D,p} + p^{D}_{j,t} + v_{b^D,j,t} \cdot G_{l|s(l)=b^D,j} \nonumber  =0,  
\end{flalign}
where $G_{l,j}$ is conductance of line $l$. If bus $b^D$ does not contain generators, then $\sum_{i \in \mathcal{I}^{D}_{b^D}} g_{i,j,t}^{D,p}\! =\! 0$, if bus $b^D$ is not a root bus ($b^D \notin \mathcal{J}^{T}$), then interface active power exchange levels are zero $p^{D}_{j,t} = 0$, and if bus $b^D$ does not contain load, then $L_{b^D,j,t}^{D,p} = 0$. Because of the radial topology, the summation of power flows is performed with respect to lines whereby $b^T$ is a ``sending" bus because any node can have several children nodes and at most one parent node.  Thus, any node is able to receive power from only one parent, although with a loss of $a_{l,j,t} \cdot G_{l,j}$. Nodal power flow for reactive power is similarly introduced: 
\begin{flalign}
&   \sum_{l|s(l)=b^D} f_{l,j,t}^{D,q} -   \left(f_{l,j,t}^{D,q} - a_{l,j,t} \cdot X_{l,j} \right)_{l|r(l)=b^D} -\label{dso_eq9} \\ & \hspace{0.2cm} \sum_{i \in \mathcal{I}^{D}_{b^D}} g_{i,j,t}^{D,q} +   L_{b^D,j,t}^{D,q} + q^{D}_{j,t} + v_{b^D,j,t} \cdot B_{l|s(l)=b^D} \nonumber  =0,  
\end{flalign}
where $B_{l,j}$ is susceptance of line $l$. 
\subsection{Coordinated TSO-DSO Model} \label{sec:coordinated}
Operating decisions of the TSO and multiple DSOs are coordinated by solving the following optimization problem:
\begin{flalign}
& \min_{\substack {\textcolor{blue}{\mathbf{F}^T}, \mathbf{G}^T,\mathbf{P}^T, \mathbf{Q}^T, \textcolor{blue}{\mathbf{V}^T}, \textcolor{blue}{\mathbf{X}^T}  \\  \textcolor{blue}{a^D_j, f^D_j}, g^D_j,p^D_j,q^D_j,\textcolor{blue}{v^D_j}}}
\begin{Bmatrix} 
O^T  \left(\mathbf{G}^T, \mathbf{P}^T, \mathbf{Q}^T \right) + \\[6pt] \displaystyle\sum_{j \in \mathcal{J}^{T}}  O^D_j  \left(g^D_j,p^D_j,q^D_j; \mathbf{\Lambda}_{j} \right)
\end{Bmatrix}, \label{coord_eq2} \\
& \hspace{3mm} s.t., \eqref{tso_eq2} - \eqref{tso_eq13}, \eqref{dso_eq2} - \eqref{dso_eq9}.  \label{coord_eq3} 
\end{flalign}
\noindent Moreover, the following interface power flow constraints ensure that the amount bought at the TSO level equals to the power sold at the DSO level, and vice versa. 

\noindent \textbf{Interface Power Flow Constraints.}
\begin{flalign}
& \mathbf{P}^{D} = \mathbf{P}^T, \mathbf{Q}^{D} = \mathbf{Q}^T,  \label{coord_eq4}
\end{flalign}
\noindent where $\mathbf{P}^{D}$ and $\mathbf{Q}^{D}$ are vectors that consist of $\{p^D_j\}$ and $\{q^D_j\}$, respectively.

 
Moreover, as pointed out by \cite{Silva}, levels of flexibility provided by DSOs is impacted by feasible ranges of power, and the $PQ$ interdependence needs to be captured at the TSO-DSO interface. Therefore, the following interface power exchange limit constraints are also imposed:
\begin{flalign}
\sqrt{\big(p^{T}_{j,t}\big)^2 \!+\! \big(q^{T}_{j,t}\big)^2} \leq \!\overline{PQ}_{j}^{T}, \sqrt{\big(p^{D}_{j,t}\big)^2\! +\! \big(q^{D}_{j,t}\big)^2} \leq \!\overline{PQ}_{j}^{D}.
\label{tsodso_eq1} 
\end{flalign} 
The problem \eqref{coord_eq2}-\eqref{tsodso_eq1} is mixed-integer nonlinear programming (MINLP) problem, which is difficult in view of the combinatorial complexity introduced by binary unit commitment variables within generation capacity \eqref{tso_eq2}-\eqref{tso_eq2b} and ramp-rate constraints \eqref{tso_eq3}-\eqref{tso_eq4a}, and the non-linearity introduced by AC power flow \eqref{tso_eq10}-\eqref{tso_eq11}, voltage restrictions \eqref{tso_eq12a}-\eqref{tso_eq12b}, transmission capacity constraints \eqref{tso_eq13} and the interface power flow restrictions \eqref{tsodso_eq1}. Moreover, cross-product terms within AC power flows \eqref{tso_eq10}-\eqref{tso_eq11} and voltage restrictions \eqref{tso_eq12a} contribute to the non-convexity of the problem.  
   
    \textcolor{blue}{Through optimization, the cost of the entire system is minimized. In this Section, unlike in the reference \cite{Cadre}, we are not considering conflicting objectives. Rather, we are interested in the development of the solution methodology to solve TSO-DSO problems efficiently in the following Section III. It is important to note that the proposed methodology is general and does need modifications to accommodate other objective functions.}
    \textcolor{blue}{In this paper, it is also assumed that DERs are located at the DSO level, so the model is ``DSO-managed'' following the definition within \cite{Givi}.}

\IEEEpeerreviewmaketitle

\section{Solution Methodology}\label{sec:solution_technique}

To resolve the above difficulties, a novel solution methodology is developed based on the recent Surrogate Lagrangian Relaxation method \cite{SLR, SAVLR}. The main ideas behind the new method are the decomposition into manageable subproblems, linearization of resulting subproblems, coordination of subproblem solutions through the update of Lagrangian multipliers, and penalization of constraint violations as well as ``$l_1-$ proximal'' terms to ensure feasibility.  Moreover, the convergence of the resulting method is provided. In subsection IV.A, the TSO-DSO coordination problem \eqref{coord_eq2}-\eqref{tsodso_eq1} is decomposed into TSO and DSO subproblems after relaxing interface power flow constraints \eqref{coord_eq4}.  To obtain location marginal prices $\mathbf{\Lambda}_j$, power flow balance constraints \eqref{tso_eq7} and \eqref{tso_eq7a} are also relaxed. In subsection IV.B, to overcome non-linearity difficulties at the TSO level, which arise because of cross-products within AC power flows \eqref{tso_eq10} and \eqref{tso_eq11}, voltage restrictions \eqref{tso_eq12a}-\eqref{tso_eq12b}, transmission capacity constraints \eqref{tso_eq13}, and interface power exchange limit constraints \eqref{tsodso_eq1}, dynamic linearization is then developed.  \textcolor{blue}{Moreover, to overcome combinatorial complexity, the decomposition into ``zonal'' subproblems within the TSO is developed.}  The entire algorithm is presented in subsection IV.C.  

\subsection{Surrogate Absolute-Value Lagrangian Relaxation}

\noindent \textbf{Relaxed Problem.} After relaxing nodal flow balance \eqref{tso_eq7}-\eqref{tso_eq7a} and interface power exchange constraints \eqref{coord_eq4}, and after penalizing their violations, the relaxed problem becomes:
\vspace{-2mm}
\begin{flalign}
& \nonumber \min_{\substack {\textcolor{blue}{\mathbf{F}^T}, \mathbf{G}^T,\mathbf{P}^T, \\ \mathbf{Q}^T, \textcolor{blue}{\mathbf{V}^T}, \textcolor{blue}{\mathbf{X}^T}  \\  \textcolor{blue}{a^D_j, f^D_j}, g^D_j,p^D_j,q^D_j,\textcolor{blue}{v^D_j}}}  L_c(\mathbf{G}^T,\mathbf{P}^T,\mathbf{Q}^T, g^D_j,p^D_j,q^D_j;\mathbf{\Lambda})  =  \\  
& \min_{\substack {\textcolor{blue}{\mathbf{F}^T}, \label{relaxedproblem} \mathbf{G}^T,\mathbf{P}^T, \\ \mathbf{Q}^T, \textcolor{blue}{\mathbf{V}^T}, \textcolor{blue}{\mathbf{X}^T}  \\  \textcolor{blue}{a^D_j, f^D_j}, g^D_j,p^D_j,q^D_j,\textcolor{blue}{v^D_j}}}  
\begin{Bmatrix}
O^T(\mathbf{G}^T,\mathbf{P}^T,\mathbf{Q}^T)   \\[7pt]+ \displaystyle\sum_{j \in \mathcal{J}^{T}} O^D_j  \left(g^D_j,p^D_j,q^D_j; \mathbf{\Lambda}_j \right) \\+ \mathbf{\Lambda}  \cdot \mathbf{R} +  c \cdot \big\| \mathbf{R}\big\|_1 
\end{Bmatrix},
\\ 
\nonumber & s.t., \eqref{tso_eq2} - \eqref{tso_eq4a},  \eqref{tso_eq10} - \eqref{tso_eq13}, \eqref{dso_eq2} - \eqref{dso_eq9}
\end{flalign}
\noindent where $\mathbf{\Lambda}_j = \big(\mathbf{\Lambda}^P_j, \mathbf{\Lambda}^Q_j\big)$ with $\mathbf{\Lambda}^P_j$ denoting multipliers that relax active power flow balance constraints \eqref{tso_eq7} and $\mathbf{\Lambda}^Q_j$ denoting multipliers that relax reactive power flow balance constraints \eqref{tso_eq7a} at root bus $j$. The vector $\mathbf{\Lambda}=(\mathbf{\Lambda}^P, \mathbf{\Lambda}^Q, \mathbf{\Psi}^P, \mathbf{\Psi}^Q)$ denotes a vector of Lagrangian multipliers \textcolor{blue}{with $(\mathbf{\Psi}^P, \mathbf{\Psi}^Q)$ relaxing constraints \eqref{coord_eq4}}, $\mathbf{R}= (\mathbf{B}^P , \mathbf{B}^Q , \mathbf{P}^{D} - \mathbf{P}^T , \mathbf{Q}^{D} - \mathbf{Q}^T)^T$ denotes a vector of constraint violations: $\mathbf{B}^P$ and $\mathbf{B}^Q$ denote violation levels of power flow balance constraints \eqref{tso_eq7}-\eqref{tso_eq7a}, and $\mathbf{P}^{D} - \mathbf{P}^T$ and $\mathbf{Q}^{D} - \mathbf{Q}^T$ are violation levels of interface power exchange constraints \eqref{coord_eq4}. The relaxed problem \eqref{relaxedproblem} is separated into TSO and DSO subproblems as explained next. 

\noindent \textbf{TSO ``Absolute-Value" Relaxed Subproblem.} \textcolor{blue}{At iteration $k$, the TSO's subproblem can be written as follows:}
\begin{flalign}
& \min_{\textcolor{blue}{\mathbf{F}^T}, \mathbf{G}^T,\mathbf{P}^T,\mathbf{Q}^T, \textcolor{blue}{\mathbf{V}^T}, \textcolor{blue}{\mathbf{X}^T}}  \bigg\{L^T_{c^{k-1}}(\mathbf{G}^T,\mathbf{P}^T,\mathbf{Q}^T;\mathbf{\Lambda}^{k-1}) \bigg\}= \nonumber \\
& \min_{\substack{\textcolor{blue}{\mathbf{F}^T}, \mathbf{G}^T,\mathbf{P}^T,\\\mathbf{Q}^T, \textcolor{blue}{\mathbf{V}^T}, \textcolor{blue}{\mathbf{X}^T}}}  
\begin{Bmatrix}
O^T(\mathbf{G}^T,\mathbf{P}^T,\mathbf{Q}^T) + \\[6pt] \mathbf{\Lambda^{k-1}}   \cdot  \mathbf{R}^{k-1} +  c^{k-1}  \cdot  \big\|  \mathbf{R}^{k-1}\big\|_1  \label{tso_sub}
\end{Bmatrix},
\\ & s.t., \eqref{tso_eq2} - \eqref{tso_eq4a},  \eqref{tso_eq10} - \eqref{tso_eq13}, \nonumber
\end{flalign}


\noindent where $\mathbf{R}^{k-1}= \begin{pmatrix} \mathbf{B}^P , \mathbf{B}^Q , (\mathbf{P}^{D})^{k-1} - \mathbf{P}^T , (\mathbf{Q}^{D})^{k-1} - \mathbf{Q}^T\end{pmatrix}^T$ is defined in a similar way to $\mathbf{R}$ above, except that interface power exchange values $(\mathbf{P}^{D})^{k-1}$ and $(\mathbf{Q}^{D})^{k-1}$ from the DSO are not decision variables, rather, they are fixed at most recent values obtained up to the iteration $k-1$.  \textcolor{blue}{Within the surrogate framework \cite{SLR}, the exact optimum of a subproblem is not required to update multipliers, rather, a satisfaction of the surrogate optimality condition is sufficient, which in terms of the problem under consideration can be written as: 
\begin{flalign}
&  L_{c^{k-1}}((\mathbf{G}^{T})^k,(\mathbf{P}^{T})^k,(\mathbf{Q}^{T})^k;\mathbf{\Lambda}^{k-1}) < \label{SOC1}\\
& L_{c^{k-1}}((\mathbf{G}^{T})^{k-1},(\mathbf{P}^{T})^{k-1},(\mathbf{Q}^{T})^{k-1};\mathbf{\Lambda}^{k-1}), \nonumber
\end{flalign}}
\noindent \textcolor{blue}{where $(\mathbf{G}^{T})^k,(\mathbf{P}^{T})^k,(\mathbf{Q}^{T})^k$ is a feasible solution to subproblem \eqref{tso_sub}. Subsequently, values $(\mathbf{P}^{T})^k$ and $(\mathbf{Q}^{T})^k$ are passed on to the DSO subproblem and are used to update multipliers together with $(\mathbf{B}^{P})^k$ and $(\mathbf{B}^{Q})^k$ as discussed next.}

\noindent \textbf{Multiplier Update.} The multipliers are updated as follows: 
 \begin{flalign}
 \mathbf{\Lambda}^{k}  
=
\mathbf{\Lambda}^{k-1}  
+ s^k \cdot \mathbf{R}^k, \label{multiplierupdate}
 \end{flalign}
 \noindent where $\mathbf{R}^{k}= ((\mathbf{B}^{P})^k, (\mathbf{B}^{Q})^k, (\mathbf{P}^{D})^{k-1}  - (\mathbf{P}^{T})^k, (\mathbf{Q}^{D})^{k-1} - (\mathbf{Q}^{T})^k)^T.$ 
 
 \noindent \textbf{Stepsize Update.} Following \cite{SAVLR}, the stepsize is updated in the following way:
\begin{flalign}
& s^{k}= \alpha^{k-1} \cdot s^{k-1} \cdot \label{eq12}
\frac{\big\| \mathbf{R}^{k-1} \big\|_2}{\big\| \mathbf{R}^{k} \big\|_2}.
\end{flalign}

\noindent  where $\alpha^{k-1}$ is a step-sizing parameter
  \begin{flalign}
& \alpha^{k-1} = 1-\frac{1}{M \cdot k^{1-\frac{1}{k^r}}}, M>1, r>0. \label{eq12a}
 \end{flalign}
 
\noindent \textbf{Penalty Coefficient Update.} In the beginning of the iterative process, the penalty coefficient $c^k$ increases by a predetermined constant $\beta > 1$:
  \begin{flalign}
& c^{k} =  c^{k-1} \cdot \beta. \label{penaltyincrease}
 \end{flalign}
 
\noindent The intent is to increase the value of $c^k$ until the norm of constraint violations reduces to zero and a feasible solution is obtained, after which the penalty coefficient is decreased per
  \begin{flalign}
& c^{k} = c^{k-1} \cdot \beta^{-1} \label{penaltyreduce}
 \end{flalign}
 
\noindent Subsequently, the penalty coefficient is not increased.

\noindent \textbf{DSO ``Absolute-Value" Relaxed Subproblem.} \textcolor{blue}{At iteration $k+1$, the DSOs' subproblem can be written as follows:}

\begin{flalign}
& \min_{\substack{\textcolor{blue}{a^D_j, f^D_j}, g^D_j\!,\\p^D_j,  q^D_j,\textcolor{blue}{v^D_j}}} 
\begin{Bmatrix} 
O^D_j \left(g^D_j,p^D_j,q^D_j; \mathbf{\Lambda}_j^k \right) + \\[6pt]
((\mathbf{\Psi}^{P})^k\!, (\mathbf{\Psi}^{Q})^k) \!\cdot\! (\mathbf{P}^{D} \!\!- \!(\mathbf{P}^{T})^k, \mathbf{Q}^{D}\!\! -\! (\mathbf{Q}^{T})^k)^T 
\\[6pt] + \; {c^k} \cdot \big\|\mathbf{P}^{D} - (\mathbf{P}^{T})^k, \mathbf{Q}^{D} - (\mathbf{Q}^{T})^k\big\|_1  \label{dso_sub}
\end{Bmatrix} \\ 
& \hspace{4mm} s.t., \eqref{dso_eq2} - \eqref{dso_eq9}. \nonumber 
\end{flalign}

\noindent The DSOs' decisions to buy/sell power from the TSO are affected not only by the nodal shadow prices $\mathbf{\Lambda}_j^{k}$ but also by multipliers $((\mathbf{\Psi}^P)^{k} \!, (\mathbf{\Psi}^Q)^{k})$ - the shadow prices that correspond to interface power exchange constraints \eqref{coord_eq4}. \textcolor{blue}{For DSO subproblems, the surrogate optimality condition becomes: 
\begin{flalign}
&  L_{c^{k}}((g^{D}_j)^{k+1},(p^{D}_j)^{k+1},(q^{D}_j)^{k+1};\mathbf{\Lambda}^{k}) < \label{SOC2}\\
& L_{c^{k}}((g^{D}_j)^{k-1},(p^{D}_j)^{k-1},(q^{D}_j)^{k-1};\mathbf{\Lambda}^{k}). \nonumber
\end{flalign}}
\noindent \textcolor{blue}{Within the surrogate Lagrangian relaxation framework, it is not necessary to solve all DSO subproblems at the same time to satisfy the above condition \eqref{SOC2} in order to update multipliers.}\footnote{\textcolor{blue}{For simplicity, here and for the rest of the paper, it is assumed that all DSO subproblems are solved at the same time.}} \textcolor{blue}{Following \eqref{multiplierupdate}, \eqref{eq12} and \eqref{eq12a}, the multipliers and stepsizes are updated in the following way:}

\noindent \textbf{Multiplier Update.} 
 \begin{flalign}
 \mathbf{\Lambda}^{k+1}  
=
\mathbf{\Lambda}^{k}  
+ s^{k+1} \cdot \mathbf{R}^{k+1}, \label{multiplierupdate1}
 \end{flalign}
 \noindent where $\mathbf{R}^{k+1} = ((\mathbf{B}^{P})^k, (\mathbf{B}^{Q})^k, (\mathbf{P}^{D})^{k+1} - (\mathbf{P}^{T})^k, (\mathbf{Q}^{D})^{k+1} - (\mathbf{Q}^{T})^k)^T.$

 \noindent \textbf{Stepsize Update.} 
\begin{flalign}
& s^{k+1}= \alpha^{k} \cdot s^{k} \cdot 
\frac{\big\| \mathbf{R}^{k} \big\|_2}{\big\| \mathbf{R}^{k+1} \big\|_2}, \label{eq12b}
\end{flalign}

\noindent  where $\alpha^{k}$ is
  \begin{flalign}
& \alpha^{k} = 1-\frac{1}{M \cdot {(k+1)}^{1-\frac{1}{(k+1)^r}}}, M>1, r>0. \label{eq12c}
 \end{flalign}

\textcolor{blue}{Theoretical convergence of the method is provided next.}


\noindent \textcolor{blue}{\textbf{Theorem 1.} If the surrogate optimality conditions \eqref{SOC1} and \eqref{SOC2} are satisfied, then under the stepsizing conditions \eqref{eq12}-\eqref{eq12a} and \eqref{eq12b}-\eqref{eq12c}, the Lagrangian multipliers converge to their optimal values.}

\noindent \textcolor{blue}{\textbf{Proof:} The proof is based on Theorem 2.1 from \cite[p. 180]{SLR} and on Theorem 1 from \cite[p. 535]{SAVLR}. The difference from \cite{SLR} is that because of the penalty terms based on interface power flow constraints \eqref{coord_eq4}, TSO and DSO subproblem solutions are not independent. This dependence issue was overcome in Theorem 1 from \cite[p. 535]{SAVLR} stating that while the surrogate optimality condition is satisfied, multipliers converge. While the proof of \cite{SAVLR} is for MILP problems, there is no requirement of linearity, and Theorem 1 from \cite[p. 535]{SAVLR} is applicable here for the MINLP problem as well. \qedsymbol} 

 \subsection{Practical Considerations of the Method}

In practical implementations, the following considerations
are important. 1. Voltage restrictions \eqref{tso_eq12a}-\eqref{tso_eq12b}, transmission capacity constraints \eqref{tso_eq13} and interface power exchange limits \eqref{tsodso_eq1} are nonlinear, they need to be appropriately linearized while maintaining convergence and feasibility. 2. Because of binary commitment decision variables, the TSO problem is combinatorial and further decomposition is needed. These considerations are discussed next.
 
 \noindent \textbf{Non-linearity of AC Power Flow and Voltage Restriction Constraints at TSO.}  
 
 1. \textbf{Linearization of Cross-Product Terms.} AC power flow constraints contain cross-products of voltages at neighboring buses. To linearize these constraints, voltages at one of the buses $(s(l)$ or $r(l))$ are fixed, thereby rendering the constraint linear.   Assuming that voltages at ``sending" buses are fixed at values $v_{s(l),t}^{T,k-1}$ obtained at previous iteration $k-1$, the constraint becomes:   
\begin{flalign}
& \widehat{f}_{l,t}^{T,p}= v_{s(l),t}^{T,k-1} \!\cdot \!\begin{pmatrix} g_{s(l),r(l)}  \;  \mbox{-} b_{s(l),r(l)} \\
b_{s(l),r(l)} \;\;  g_{s(l),r(l)}  \end{pmatrix} \!\cdot\! \left( v_{r(l),t}^T \right)^T. \label{linearized1}
\end{flalign}

The issue with this approach is that a subset of ``receiving" buses does not generally equal to the entire set of buses. As a result, some of the voltages are not updated.  To overcome this issue, voltages at ``sending'' buses are fixed, and voltages at ``receiving" buses are fixed independently based on \eqref{tso_eq10} and the average of the resulting power flows is taken as:
\begin{flalign}
& \widehat{f}_{l,t}^{T,p}= \frac{1}{2} v_{s(l),t}^{T,k-1} \!\cdot \!\begin{pmatrix} g_{s(l),r(l)}  \;  \mbox{-} b_{s(l),r(l)} \\
b_{s(l),r(l)} \;\;  g_{s(l),r(l)}  \end{pmatrix} \!\cdot\! \left( v_{r(l),t}^T \right)^T + \nonumber \\
& \hspace{11.5mm} \frac{1}{2} v_{s(l),t}^T \!\cdot\! \begin{pmatrix} g_{s(l),r(l)}  \; \mbox{-} b_{s(l),r(l)} \\
b_{s(l),r(l)}  \;\; g_{s(l),r(l)}  \end{pmatrix} \!\cdot\! \left( v_{r(l),t}^{T,k-1} \right)^T\!. \label{linearized2}
\end{flalign}
\noindent In doing so, all voltages are updated at every iteration. Reactive power flows are linearized in the same way. 

In the following, linearization of voltage restrictions \eqref{tso_eq12b}, transmission capacity constraints \eqref{tso_eq13} and interface power flow restrictions \eqref{tsodso_eq1} is performed. Since voltage restrictions \eqref{tso_eq12a} delineate non-convex set, their lineaization is treated separately afterward. 

2. \textbf{Linearization of \eqref{tso_eq12b}, \eqref{tso_eq13} and \eqref{tsodso_eq1}}. Since the above constraints are similar in their structure, the lineaization is demonstrated by using constraints \eqref{tso_eq12b}. The linearization is performed in two steps. In Step 1, equation \eqref{tso_eq12b} is squared to eliminate the non-linearity introduced by the square root and thus can equivalently be rewritten as: 
 \begin{flalign}
& \big(v_{b^T,t}^{T,Re}\big)^2+\big(v_{b^T,t}^{T,Im}\big)^2 \!\leq \left(\overline{v}_{b^T}^T\right)^2\!, \label{linearized3}
\end{flalign}
In Step 2, squared terms within \eqref{linearized3} are then linearized in the following way: 
 \begin{flalign}
& v_{b^T,t}^{T,Re,k-1} \cdot v_{b^T,t}^{T,Re}+v_{b^T,t}^{T,Im,k-1} \cdot v_{b^T,t}^{T,Im} \!\leq \left(\overline{v}_{b^T}^T\right)^2\!. \label{linearized4}
\end{flalign}
Linearization of \eqref{tso_eq13} and \eqref{tsodso_eq1} is performed by using exactly the same procedure.

3. \textbf{Linearization of \eqref{tso_eq12a}}. The first two steps of linearization follow those described above. Additionally, to avoid possible infeasibility, the following ``soft" constraint with the penalty variable $v^{T,pen}_{b^T,t}$ is enforced following \cite{ABB} as: 
\begin{flalign}
& \left(\underline{v}_{b^T}^T\right)^2 \! - \! v^{T,pen}_{b^T,t} \!\leq\! v_{b^T,t}^{T,Re,k-1} \!\cdot \!v_{b^T,t}^{T,Re}\!+\!v_{b^T,t}^{T,Im,k-1} \!\cdot \!v_{b^T,t}^{T,Im}. \label{linearized5}
\end{flalign}

To enforce feasibility of \eqref{linearized5},  $v^{T,pen}_{b^T,t}$ are penalized.  To avoid getting trapped at a local minimum, which may happen when penalties are high, a novel ``flexible penalization" is introduced by using Lagrangian multipliers in a non-conventional way without relaxing \eqref{linearized5}. The Lagrangian multipliers increase only when the ``soft" constraints are violated $(v^{T,pen}_{b^T,t} > 0)$, and decrease when satisfied: $(v^{T,pen}_{b^T,t} = 0)$ and $\left(\underline{v}_{b^T}^T\right)^2 < v_{b^T,t}^{T,Re,k-1} \!\cdot \!v_{b^T,t}^{T,Re}\!+\!v_{b^T,t}^{T,Im,k-1} \!\cdot \!v_{b^T,t}^{T,Im}$. When constraints are satisfied, the associated multipliers become zero. To discourage future violations of these constraints, violations $v^{T,pen}_{b^T,t}$ are also penalized by using $c^k$ as will be shown below.

To ensure that solutions satisfying constraints \eqref{linearized2}, \eqref{linearized4} and \eqref{linearized5} satisfy the original constraints \eqref{tso_eq12a}-\eqref{tso_eq12b}, \eqref{tso_eq13} and \eqref{tsodso_eq1},
 proximal-like terms $\big\| \mathbf{V}^T \!\!-\! (\mathbf{V}^{T})^{k-1} \big\|_1$, which capture the deviations of voltages from previously obtained values are first introduced and then penalized by \textcolor{blue}{$c^k_p$ (subscript ``$p$'' is for ``proximal''). The intent to have a separate penalty coefficient, lower in value as compared to $c^k$, is to avoid solutions getting trapped at previously obtained values.} Here $\mathbf{V}^T$ is a vector of voltages $\big(\{v_{b^T,t}^{T,Re}\},\{v_{b^T,t}^{T,Im}\}\big)$. Following the same logic, to satisfy \eqref{tso_eq13} and \eqref{tsodso_eq1}, the following terms are introduced $\big\| \mathbf{\widehat{F}}^T \!\!-\! (\mathbf{\widehat{F}}^{T})^{k-1} \big\|_1$ and $\big\| \mathbf{PQ}^T \!\!-\! (\mathbf{PQ}^{T})^{k-1} \big\|_1$, where $\mathbf{\widehat{F}}^T$ is a vector of AC power flows $\big(\{\widehat{f}_{l,t}^{T,p}\},\{\widehat{f}_{l,t}^{T,q}\}\big)$ and $\mathbf{PQ}^T$ is a vector of interface power exchange flows $\big(\{p^{T}_{j,t}\},\{q^{T}_{j,t}\}\big)$.


\noindent \textbf{Linearized TSO Subproblem.} 
The resulting TSO subproblem then becomes: 
 \begin{flalign}
& \min_{\substack  {\textcolor{blue}{\mathbf{\widehat{F}}^T}, \mathbf{G}^T,\mathbf{P}^T, \\ \mathbf{Q}^T, \textcolor{blue}{\mathbf{V}^T}, \textcolor{blue}{\mathbf{X}^T}}} \bigg\{O^{T,Sub}(\textcolor{blue}{\mathbf{\widehat{F}}^T}, \mathbf{G}^T,\mathbf{P}^T,\mathbf{Q}^T,\textcolor{blue}{\mathbf{V}^T})\bigg\} = \nonumber \\ 
&\min_{\substack  {\textcolor{blue}{\mathbf{\widehat{F}}^T}, \mathbf{G}^T,\mathbf{P}^T, \\ \mathbf{Q}^T, \textcolor{blue}{\mathbf{V}^T}, \textcolor{blue}{\mathbf{X}^T}}}  
\begin{Bmatrix} L^T_{c^{k-1}}(\mathbf{G}^T,\mathbf{P}^T,\mathbf{Q}^T;\mathbf{\Lambda}^{k-1}) +\\[3pt]  c^{k-1}_p \cdot  \big\| \mathbf{V}^T - (\mathbf{V}^{T})^{k-1} \big\|_1  + \\[3pt] c^{k-1}_p \cdot \big\| \mathbf{\widehat{F}}^T - (\mathbf{\widehat{F}}^{T})^{k-1} \big\|_1  + \\[3pt] c^{k-1}_p \cdot \big\| \mathbf{PQ}^T \!-\! (\mathbf{PQ}^{T})^{k-1} \big\|_1 + \\[3pt] c^{k-1}\!\!  \cdot \!\!\!\!\!\!\!\!\! \displaystyle\sum_{t \in \mathcal{T}, b^T \in \mathcal{B}^{T}} \!\!\!\!\!\!\!v^{T,pen}_{b^T\!\!,t} + \!\!\!\!\!\!\!\!\displaystyle\sum_{t \in \mathcal{T}, b^T \in \mathcal{B}^{T}} \!\!\!\!\!\!\!\! \lambda^{T,pen}_{b^T\!\!, t} \!\cdot\! v^{T,pen}_{b^T\!\!,t}   
\end{Bmatrix}, \label{linearizedtsosubproblemconstraints}
\\
& \hspace{0mm} s.t., \eqref{tso_eq2} - \eqref{tso_eq4a},  \eqref{tso_eq10} - \eqref{tso_eq13}. \nonumber 
\end{flalign}
Flexibility and versatility of Lagrangian multipliers not only allows to coordinate subsystems in the inter-subproblem way but also to resolve non-convexity in the intra-subproblem manner to avoid local minima which may be caused by \eqref{tso_eq12a}. The multiplier update for \eqref{linearizedtsosubproblemconstraints} is operationalized by appending constraints violations $\mathbf{R}$ by $v^{T,pen}_{b^T\!\!,t}$ and multipliers $\mathbf{\Lambda}$ by $\lambda^{T,pen}_{b^T\!\!,t}$, and by following the multiplier updating procedure described in \eqref{multiplierupdate} with projections of negative values of $\lambda^{T,pen}_{b^T\!\!,t}$ onto a subspace $\{\lambda | \lambda \ge 0\}$. Piece-wise linear $l_1$-norms within \eqref{linearizedtsosubproblemconstraints} are linearized following standard procedures \cite{SALR, SAVLR}.

\noindent \textcolor{blue}{\textbf{Decomposition into ``Zonal'' Subproblems}.
The problem \eqref{linearizedtsosubproblemconstraints}, while easier to solve as compared to \eqref{coord_eq2}-\eqref{tsodso_eq1} because of independence of DSO subproblems and because of linearity, may still be difficult to solve because of combinatorial complexity - the problem \eqref{linearizedtsosubproblemconstraints} contains all the binary ``unit commitment'' $x_{i,t}$ decision variables. This is a general difficulty behind such methods as Benders decomposition whereby all binary variables are within the Master problem. To overcome the complexity difficulty, the TSO subproblem is decomposed into ``zonal subproblems:''
 \begin{flalign}
& \min_{\substack{\{\mathbf{G}^T\!,\mathbf{P}^T\!,\mathbf{Q}^T\!, \textcolor{blue}{\mathbf{V}^T}\!, \textcolor{blue}{\mathbf{X}^T}\} \in \mathbf{Z} \\ \mathbf{L}:  \{\mathbf{S},\mathbf{R}\}\in \mathbf{Z} }}  \bigg\{O^{T,Sub}_Z(\mathbf{\widehat{F}}^T\!, \mathbf{G}^T\!,\mathbf{P}^T\!,\mathbf{Q}^T\!, \mathbf{V}^T)\bigg\} \label{zonalsubproblem} \\
& \hspace{0mm} s.t., \eqref{tso_eq2} - \eqref{tso_eq4a},  \eqref{tso_eq10} - \eqref{tso_eq13}. \nonumber 
\end{flalign}
\noindent where $\{\mathbf{S},\mathbf{R}\}$ is a set of sending and receiving buses belonging to zone $\mathbf{Z}$ and $\{\mathbf{L}:  \{\mathbf{S},\mathbf{R}\}\in \mathbf{Z}\}$ is a set of power lines with either sending or receiving buses (or both) belonging to zone $\mathbf{Z}.$   In a sense, if neither a sending nor a receiving bus belongs to zone $\mathbf{Z},$ then all voltages within \eqref{tso_eq10}-\eqref{tso_eq11} are fixed at the most recently obtained values; otherwise, power flows follow \eqref{linearized2}. One possible partitioning of a transmission system is shown in Figure 1 based on the IEEE 118-bus system.
\begin{figure}[]
  \centering
    \includegraphics[trim=60 375 60 90, width=1\linewidth, scale=0.3]{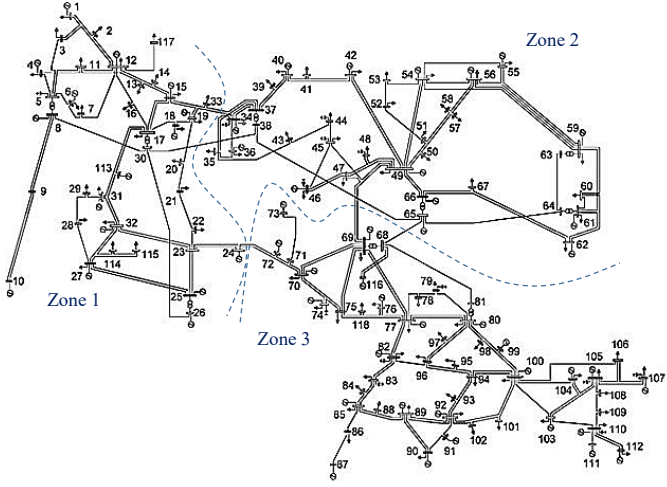}
    \caption{Topology of the TSO system for Case Study 3 based on IEEE 118-bus system.}
    \label{fig_ex1_topology}
\end{figure}
}

\noindent \textcolor{blue}{\textbf{Theorem 2.} If solutions to the linearized TSO ``zonal'' subproblems \eqref{zonalsubproblem} satisfy the surrogate optimality condition \eqref{SOC1} and solutions to DSO subproblems \eqref{dso_sub} satisfy \eqref{SOC2}, then under the stepsizing conditions \eqref{eq12}-\eqref{eq12a} and \eqref{eq12b}-\eqref{eq12c}, the Lagrangian multipliers converge to their optimal values.}

\noindent \textcolor{blue}{\textbf{Proof:} The proof follows that of Theorem 1. \qedsymbol}


\noindent \textcolor{blue}{\textbf{Feasibility.} As multipliers approach their optimal values and as penalty coefficients $c^k$ decrease, the violations of relaxed constraints $\mathbf{R^k}$ and $v^{T,pen}_{b^T,t}$ decrease. However, zero constraint violations do not imply feasibility because power flows $\widehat{f}_{l,t}^{T,p}$ and $\widehat{f}_{l,t}^{T,q}$ are the linearized versions of the actual power flows $f_{l,t}^{T,p}$ and $f_{l,t}^{T,q}$. To ensure that $\widehat{f}_{l,t}^{T,p} \rightarrow f_{l,t}^{T,p}$ and $\widehat{f}_{l,t}^{T,q} \rightarrow f_{l,t}^{T,q}$, penalty coefficients $c^k_p$ increase in a manner similar to \eqref{penaltyincrease} as: 
  \begin{flalign}
& c^{k}_p =  c^{k-1}_p \cdot \beta_p. \label{penaltyincrease_p}
 \end{flalign}
\noindent The intent is to increase the value of $c^k_p$ until the ``$l_1-$proximal'' terms reduce to zero and a feasible solution is obtained, after which the penalty coefficient is decreased per
  \begin{flalign}
& c^{k}_p = c^{k-1}_p \cdot \beta^{-1}_p. \label{penaltyreduce_p}
 \end{flalign}
Note that when penalty coefficient $c^k$ is high, the surrogate optimality condition may not be satisfied, following Proposition 1 from \cite[p. 535]{SAVLR}. In a similar way, when $c^k_p$ is high, then the surrogate optimality condition may not be satisfied as well, so that reduction of penalty coefficients per \eqref{penaltyreduce} and \eqref{penaltyreduce_p} is justified. After constraints violations are zero, the feasible solution is obtained and after the reduction of penalty coefficients, multipliers are updated again if the corresponding TSO and DSO surrogate optimality conditions are satisfied until constraint violations and proximal terms are zero again, and the process repeats. 
When solving large-scale problems, the reduction of constraint violations and proximal terms to exactly zero may require significant CPU time, so the algorithm stops when constraint violations are less than a predetermined value $\varepsilon$ and proximal terms are less than $\varepsilon_p$.}

 \subsection{Algorithm.}
\vspace{-2mm}
 
\begin{algorithm}[h!]
\SetAlgoLined
Initialize $\mathbf{\Lambda}^{0}$, $c^0$, \textcolor{blue}{$c^0_p$} and $s^0$ \\
\While{stopping criteria are not satisfied}
{
  \textbf{1} solve TSO subproblem \eqref{linearizedtsosubproblemconstraints} using fixed values $(\mathbf{P}^{D})^{k-1}$ and $(\mathbf{Q}^{D})^{k-1}$\;
  \textbf{2}
  \eIf{the ``surrogate optimality condition'' \eqref{SOC1} is satisfied}
  {update multipliers per \eqref{multiplierupdate} and increase $c^k$; goto 3}
  {goto 1}
  \textbf{3} solve DSO subproblems \eqref{dso_sub} with $\!(\mathbf{P}^{T})^k\!$ and $\!(\mathbf{Q}^{T})^k\!$\;
  \textbf{4} 
  \eIf{the ``surrogate optimality condition'' \eqref{SOC2} is satisfied}
  {update multipliers \eqref{multiplierupdate1}; goto 1}
  {goto 1}
  \eIf{constraint violations $\mathbf{R^k}$ and $v^{T,pen}_{b^T,t}$ are less than $\varepsilon$}{
  \eIf{``$l_1-$proximal'' terms $\big\| \mathbf{V}^T \!\!-\! (\mathbf{V}^{T})^{k-1} \big\|_1$, $\big\| \mathbf{\widehat{F}}^T \!\!-\! (\mathbf{\widehat{F}}^{T})^{k-1} \big\|_1$ and $\big\| \mathbf{PQ}^T \!\!-\! (\mathbf{PQ}^{T})^{k-1} \big\|_1$ are less than $\varepsilon_p$}{calculate feasible costs based on the objective functions \eqref{tsoobjective} and \eqref{dso_obj}. Decrease $c^k_p$ and $c^k$.}{Increase $c^k_p$; goto 1;}
  }{ 
   goto 1
  }
 }
 \caption{Surrogate Lagrangian Relaxation}
\end{algorithm}

 \noindent \textbf{Stopping Criteria.} The algorithm is terminated after the stopping criteria are satisfied. The following stopping criteria may be used: CPU time limit, number of iterations, etc.

\section{Numerical Testing}

The TSO-DSO coordination approach developed in this paper is implemented in CPLEX 12.10.0.0 by using a laptop with the processor Intel® Xeon® CPU E3-1535M v6 @ 3.1-GHz and 32.00 GB of RAM.  Three case studies are considered to illustrate the important concepts: convergence of multipliers, the satisfaction of nonlinear constraints, scalability, and cost benefits for TSO and DSOs.  Within Case Study 1, a small 1-hour example with 4 buses 2 generators within TSO, and one DSO is considered to demonstrate the convergence of multipliers.  Within Case Study 2, a small 9-bus system \cite{matpower} is considered to demonstrate the efficiency of linearization and convergence of voltages to satisfy constraints \eqref{tso_eq12a}-\eqref{tso_eq12b}.  Within Case Study 3.a, 4-, 8- and 12-hour instances with 1 TSO modeled using IEEE 118-bus system \cite{matpower} and 32 DSOs modeled using IEEE 34-bus systems \cite{matpower} are considered. \textcolor{blue}{Within Case Study 3.b, 24-hour problem instance with 1 TSO modeled using IEEE 118-bus system \cite{matpower} and 4 DSOs modeled using IEEE 34-bus systems \cite{matpower} are considered. Both Case studies 3.a and 3.b are} to demonstrate the method's ability to coordinate \textcolor{blue}{a different number of DSO with different planning horizons} and the cost advantage of the coordination. 

\subsection{Case Study 1: Simple TSO-DSO Coordination Example.} \label{sec:study_illustration}

Consider a simple TSO-DSO topology with the corresponding technical characteristics such as resistance and inductance as well as generation costs, power interface exchange prices, and loads (Fig. \ref{fig_ex1_topology}): TSO consists of 4 buses, includes 2 generators (located at buses 1 and 4), and containing one root bus (bus 2) whereby one DSO is connected modeled using data for the IEEE 34-bus system. \textcolor{blue}{Within this Case Study, $s^0 = 0.01, c^0 = 10, c_p = 10^{-4}, \beta = 1.025, \beta_p = 1.01, \mathbf{\Lambda} = 0, \varepsilon = 10^{-6}$ and $\varepsilon_p = 10^{-6}$. The method converges within 90 iterations and in total takes $2.3$ sec ($1.4$ sec for the TSO and $0.9$ sec for the DSO).}  The resulting dispatch including power generation and power flow as well as values of multipliers at convergence are also shown in Fig. \ref{fig_ex1_topology} below.

\begin{figure}[!htb]
  \centering
    \includegraphics[trim=10 450 10 80, width=1\linewidth, scale=0.3]{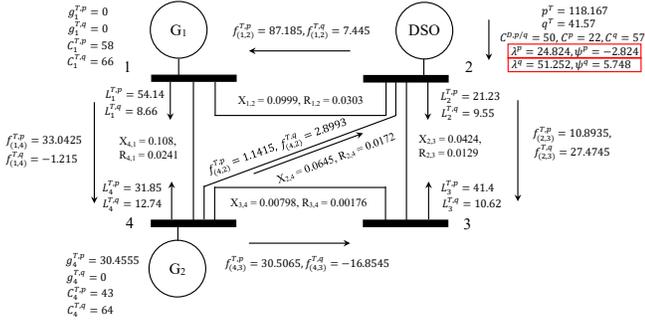}
    \caption{Topology of the TSO-DSO system for Case Study 1.}
    \label{fig_ex1_topology}
\end{figure}

  As demonstrated in Fig. \ref{fig_ex1_topology}, to serve the loads, TSO purchases power from the DSO $p^T_2=118.167, q^T_2=41.57$ and generates power using generator at bus 4: $g^{T,p}_4=30.4555$. 
  
\begin{figure}[!htb]
  \centering
    \includegraphics[trim=50 210 50 210, width=1\linewidth, scale=0.3]{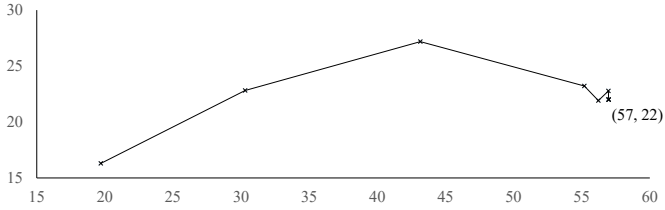}
    \caption{Convergence of $\lambda^p_2 + \psi^p_2$ and $\lambda^q_2 + \psi^q_2$ at the root bus $2$.}
    \label{fig_convergence_gap}
\end{figure}

  Fig. \ref{fig_convergence_gap} below demonstrates convergence of the multipliers at the root bus. 
  
 Multipliers $\lambda^p_2$ and $\lambda^q_2$ are shadow nodal prices, and $\psi^p_2$ and $\psi^q_2$ are shadow prices associated with interface power exchange constraints.  Total shadow prices associated with the TSO-DSO power exchange are $\lambda^p_2 + \psi^p_2$ and $\lambda^q_2 + \psi^q_2$ as shown in Fig. \ref{fig_convergence_gap}. Prices approach the values of $\$22$ for active and $\$57$ for reactive power, which are exactly equal to the DSO's interface power exchange amount bids $C^p_2$ and $C^q_2$.

To demonstrate the cost-advantage of TSO-DSO coordination, a series of $30$ Monte Carlo simulations is performed by randomly generating generation costs, as well as DSO's bids, by using a uniform distribution U[20,60]. \textcolor{blue}{While in several cases, the TSO or DSO cost improvement was negative,} on average, DSOs' total cost is 4.18\% lower and TSO's cost is 13.57\% lower as compared to DSOs and TSO cost obtained by solving DSO and TSO problems separately.    

\subsection{Case Study 2: Illustration of Efficiency of Dynamic Linearization Based on the 9-bus System.} 

Within this Case Study, consider a 9-bus system \cite{matpower}. The purpose of this example is to demonstrate that the dynamic linearization is efficient and that solutions satisfy original constraints at convergence. For illustration purposes, the satisfaction with respect to voltage restriction constraints \eqref{tso_eq12a}-\eqref{tso_eq12b} is demonstrated. \textcolor{blue}{Within this Case Study, parameters of the algorithm are the same as in Case Study 1. The method converges within 64 iterations and takes $1.5$ sec.} The results are shown in Fig. \ref{Ex22} below.  

\begin{figure}[!htb]
  \centering
    \includegraphics[trim=10 150 10 150, width=0.8\linewidth, scale=0.1]{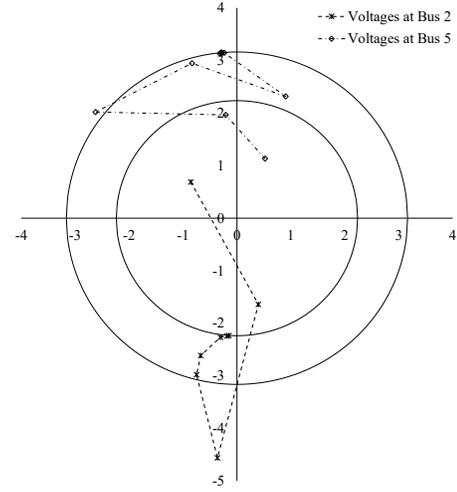}
    \caption{Convergence of voltages for Example 2 for selected buses. Real voltages are along the x-axis and imaginary voltages are along the y-axis.}
    \label{Ex22}
\end{figure}

As shown in Fig. \ref{Ex22}, the inner and outer circles are the boundaries of the feasible region corresponding to minimal and maximum voltages squared $\big(\underline{v^T_{b^T}})^2$ and $\big(\overline{v^T_{b^T}})^2$, respectively.  The feasible region is in-between the two circles. Within buses 2 and 5, initial voltages are within the inner circle and are infeasible.  Through coordination and penalization of constraint violations, voltages converge to values within the feasible region, i.e., original constraints \eqref{tso_eq12a}-\eqref{tso_eq12b} are satisfied. 

\subsection{Case Study 3: TSO-DSO Coordination with One IEEE 118-Bus System and several IEEE 34-Bus Systems} \label{sec:study_illustration}

\noindent \textcolor{blue}{Case Study 3.a:} Consider a TSO-DSO system with TSO modeled as IEEE 118-bus system \cite{matpower} (with the corresponding loads, ramp rates as well as minimum and maximum generation levels). DSOs are modeled as IEEE 34-bus systems \cite{matpower}. \textcolor{blue}{For simplicity, within each DSO, dispatcheable units located at buses 816, 840, 844, and 854 with capacities of 10 MW are considered.} \textcolor{blue}{The DSOs are identical in terms of topology, generation capacity. Within TSO and DSOs, generation costs are based on \cite{matpower} and are randomly perturbed, so that each DSO system has a unique set of parameters.} For the cost comparison purposes, the ``uncoordinated'' TSO problem\footnote{\textcolor{blue}{Within the ``uncoordinated'' case, the TSO unit commitment problem is solved separately from DSOs economic dispatch problem while ignoring interface power exchange constraints.}} 
is solved by itself using the method presented in Section IV by updating penalty coefficients per \eqref{penaltyincrease}-\eqref{penaltyreduce} and by updating multipliers per \eqref{multiplierupdate} without including $\mathbf{P}^{D,k} - \mathbf{P}^{T,k}$ and $\mathbf{Q}^{D,k} - \mathbf{Q}^{T,k}$ into $\mathbf{R}^{k}$.  The DSO problems are directly solved by using CPLEX, which can handle the types of nonlinearities that are present within the DSO problems. \textcolor{blue}{Within this Case Study, $s^0 = 0.035, c^0 = 25, c_p = 10^{-3}, \beta = 1.05, \beta_p = 1.05, \varepsilon = 10^{-2}$ and $\varepsilon_p = 10^{-2}$. For buses with generators, multipliers are initialized using generation costs, for buses without generators, multipliers are initialized at zero.} The total costs obtained are then compared with the cost obtained by the TSO-DSO coordination and the results are summarized in Table I. Relative cost improvement is shown in parentheses. 

 

     
\begin{table}[ht]
\caption{Cost improvement through TSO-DSO coordination for 1 TSO and 32 DSOs with 4-, 8- and 12-hour horizons.} 
\centering 
\begin{tabular}{c c c c c c} 
\hline\hline 
 No. & TSO cost & DSO cost & Total cost & TSO  & DSO   \\
  of hrs. & ($\times$ $\$1000$) & ($\times$ $\$1000$)  & ($\times$ $\$1000$) & time (s) & time (s)  \\[0.5ex] 
\hline 
4  &     337.65 &	 41.78 & 	 379.54 & \textcolor{blue}{153.53} & 113.26
 
 \\
   &  ($\downarrow$ 27.53\%) &   ($\downarrow$ 21.78\%) &   ($\downarrow$ 26.95\%)     \\ 
8  &   653.39 &	84.65 &	738.04 &	\textcolor{blue}{362.96} &	289.52
 
 \\ 
    &  ($\downarrow$ 26.87\%) &   ($\downarrow$ 21.01\%) &   ($\downarrow$ 26.24\%)  	  \\ 
12  &   1113.46	& 87.76	& 1201.23	& \textcolor{blue}{663.48}	& 486.15

 \\ 
    &  ($\downarrow$ 22.68\%) &   ($\downarrow$ 45.40\%) &   ($\downarrow$ 24.96\%)  	  \\ 
\hline 
\end{tabular}
\label{table:nonlin} 
\end{table}

As demonstrated in Table I, the method is capable of coordinating a large number of DSOs together with the TSO, and that the both systems benefit from the coordination. 

\noindent \textcolor{blue}{Case Study 3.b: Within this case study, the TSO and DSOs are modeled as within Case Study 3.a. Within this case study, 4 DSOs are considered and the planning horizon is 24 hours. The results are summarized in Table II.}
\textcolor{blue}{\begin{table}[h!]
\caption{Cost improvement through TSO-DSO coordination for 1 TSO and 4 DSOs with 24-hour horizons.} 
\centering 
\begin{tabular}{c c c c c} 
\hline\hline 
TSO cost & DSO cost & Total cost & TSO  & DSO   \\
  ($\times$ $\$1000$) & ($\times$ $\$1000$)  & ($\times$ $\$1000$) & time (s) & time (s)  \\[0.5ex] 
\hline 
2213.27 &	 26.87 & 	2240.15 & 2230.12 & 741.94
 \\
     ($\downarrow$ 2.49\%) &   ($\downarrow$ 26.4\%) &   ($\downarrow$ 2.87\%)     \\ 
\hline 
\end{tabular}
\label{table:nonlin} 
\end{table}}
\textcolor{blue}{Within Case Studies 3.a-b., to satisfy the stopping criteria, the number of times TSO and DSO subproblems are solved is roughly 92.} 
\textcolor{blue}{In Table II, it is demonstrated that the CPU time spent on the TSO problem is roughly $3\times$ the CPU spend on the 12-hour case. This is because subproblems, while much easier to optimize than the entire TSO problem, are still NP-hard problems and the complexity may increase non-linearly. The overall cost savings are much less ($2.87\%$ vs. $24.96\%$) as compared to the Case Study 3.a's results with 32 DSOs, which suggests that with the increasing number of DSOs involved in the coordination, the overall cost tends to decrease.}

\section{Conclusion}
This paper presents a novel formulation for coordination of TSO and multiple DSOs by considering both active and reactive power through the inclusion of nonlinear AC power flow constraints within both systems, and by considering binary UC decisions. The resulting combinatorial and the highly nonlinear problem is solved by using the decomposition and coordination ``$l_1-$ Proximal'' Surrogate Lagrangian Relaxation together with the novel dynamic linearization to satisfy nonlinear constraints. Case studies demonstrate the effectiveness of coordination through the convergence of multipliers, efficiency of linearization, and that the method is capable of coordinating a large number of DSOs, and that both the transmission and distribution systems benefit from the proposed coordination. The new solution methodology opens several directions for 1) handling non-linear constraints that delineate convex as well as non-convex feasible regions, specifically, to handle AC power flow within unit commitment problems, 2) inclusion of stochastic elements to efficiently handle uncertainties while solving unit commitment problem with meshed topology with AC flow without losses of accuracy due to approximations frequently used, and 3) game-theoretic studies to investigate \textcolor{blue}{TSO and DSOs strategies}.


%





\ifCLASSOPTIONcaptionsoff
  \newpage
\fi

\vfill



\begin{thebibliography}{1}



\bibitem{Papa2020} A. Papalexopoulos, R. Frowd, and A. Birbas, ``On the Development of Organized Nodal Local Energy Markets and a Framework for the TSO-DSO Coordination,” \textit{Electric Power Systems Research,} vol. 189, 2020.

\bibitem{YUAN2017600} 
Z. Yuan and M. R. Hesamzadeh, ``Hierarchical Coordination of TSO-DSO Economic Dispatch Considering Large-Scale Integration of Distributed Energy Resources," \textit{Applied Energy}, vol. 195, pp. 600–615, Jun. 2017.

\bibitem{Perez2014} 
I. Pérez-Arriaga and A. Bharatkumar, ``A Framework for Redesigning Distribution Network Use of System Charges Under High Penetration of Distributed Energy Resources: New Principles for New Problems," \textit{Tech. rep., CEEPR Working Paper Series no. WP-2014-006}; 2014.

\bibitem{Nikos} \textcolor{blue}{N. Savvopoulos, T. Konstantinou, and N. Hatziargyriou. ``TSO-DSO coordination in decentralized ancillary services markets,'' In \textit{IEEE 2019 International Conference on Smart Energy Systems and Technologies (SEST), pp. 1-6, 2019.}}

\bibitem{Mig} G. Migliavacca, M. Rossi, D. Six, M. Dzamarija, S. Horsmanheimo, C. Madina, I. Kockar, and J.M. Morales, ``SmartNet: A H2020 Project Analysing TSO–DSO Interaction to Enable Ancillary Services Provision from Distribution Networks," \textit{CIRED,} Glasgow, Scotland, June 2017. 

\bibitem{Mig1} G. Migliavacca, M. Rossi, H. Gerard, M. Dzamarija, S. Horsmanheimo, C. Madina, I. Kockar, G. Leclecq, M. Marroqu{\'i}n, H. Svendsen, ``TSO-DSO Coordination and Market Architectures for an Integrated Ancillary Services Acquisition: the View of the SmartNet Project," \textit{CIRED,} Paris, France, June 2018. 


\bibitem{Mig2} M. Rossi, G. Viganò, G. Migliavacca, Y. Vardanyan, R. Ebrahimi, G. Leclercq, P. Sels, M. Pavesi, T. Gueuning, J. Jimeno, N. Ruiz, G. Howorth, J. Camargo, C. Hermans, F. Spiessen, and H. Svendsen, ``Testing TSO-DSO Interaction Schemes for the Participation of Distribution Energy Resources in the Balancing Market: the SmartNet Simulator," \textit{CIRED}, Madrid, Spain, June 2019.

\bibitem{Birk} M. Birk, J. P. Chaves-Ávila, T. Gómez, and R. Tabors, ``TSO/DSO coordination in a context of distributed energy resource penetration,'' \textit{MIT Energy Initiative Reports}, 2017.

\bibitem{TD1}
H. Jain, K. Rahimi, A. Tbaileh, R. P. Broadwater, A. K. Jain and M. Dilek, ``Integrated Transmission \& Distribution System Modeling and Analysis: Need \& Advantages," \textit{In proceedings of the IEEE Power and Energy Society 2016 General Meeting}, Boston, MA, July 2016.

\bibitem{Ali}
A. Hassan and Y. Dvorkin, ``Energy Storage Siting and Sizing in Coordinated Distribution and Transmission Systems,''  \textit{IEEE Transactions on Sustainable Energy}, vol. 9, no. 4, pp. 1692-1701, Oct. 2018.

\bibitem{Grot} \textcolor{blue}{H. H. Grøttum, S. F. Bjerland, P. C. del Granado, and R. Egging, ``Modelling TSO-DSO coordination: The value of distributed flexible resources to the power system,'' In \textit{IEEE 2019 16th International Conference on the European Energy Market (EEM)}, pp. 1-6. 2019.}

\bibitem{AlSaadi} \textcolor{blue}{M. Al-Saadi, R. Pestana, R. Pastor, G. Glória, A. Egorov, F. Reis, and T. Simão, ``Survey Analysis on Existing Tools and Services for Grid and Market Stakeholders and Requirements to Improve TSO/DSO Coordination,'' In \textit{IEEE 2019 International Symposium on Systems Engineering (ISSE)}, pp. 1-7, 2019.}

\bibitem{Pilo} \textcolor{blue}{F. Pilo, G. Mauri, B. Bak-Jensen, E. Kämpf, J. Taylor, and F. Silvestro, ``Control and automation functions at the TSO and DSO interface–impact on network planning,'' \textit{CIRED-Open Access Proceedings Journal 2017}, no. 1, pp. 2188-2191, 2017.}

\bibitem{Alves} R. Alves, F. Reis, and C. Liang, ``TSOs and DSOs collaboration: The need for data exchange,'' \textit{vol. Trivent En, no. Deregulated Electricity Market Issues in South Eastern Europe,} pp. 55-62, 2015.

\bibitem{Puente} \textcolor{blue}{E.I.R. Puente, H. Gerard, and D. Six, ``A set of roles for the evolving business of electricity distribution,'' \textit{Utilities Policy}, 55, pp. 178-188. 2018.}

\bibitem{EDSO1} EDSO. ``General Guidelines for Improving TSO-DSO Cooperation." \textit{Tech. rep.,} 2015.

\bibitem{EDSO2} EDSO. ``TSO-DSO Data Management Report." \textit{Tech. rep.}; 2015.

\bibitem{Zipf} M. Zipf, and D. Möst, ``Cooperation of TSO and DSO to Provide Ancillary Services," \textit{2016 13th International conference on the European energy market (EEM),} pp. 1–6, 2016

\bibitem{Gomez} I. Gómez et al., ``Cost-Benefit Analysis of the Selected National Cases," \textit{SmartNet Project, Tech. Rep. D4.3,} 2019 

\bibitem{Mig3} M. Rossi, G. Migliavacca, G. Viganò, D. Siface, C., I. Gomez, I. Kockar, and Andrei Morch, ``TSO-DSO Coordination to Acquire Services from Distribution Grids: Simulations, Cost-benefit Analysis and Regulatory Conclusions from the SmartNet Project," \textit{Electric Power Systems Research,} vol. 189, 2020.

\bibitem{Silva} J. P. Silva, J. Sumaili, R. J. Bessa, L. Seca, M. Matos, and V. Miranda, ``The Challenges of Estimating the Impact of Distributed Energy Resources Flexibility on the TSO/DSO Boundary Node Operating Points," \textit{Computers and Operations Research}, vol. 96, pp. 294–304, 2018.

\bibitem{Silva1}  J. P. Silva, J. Sumaili, R. J. Bessa, L. Seca, M. Matos, V. Miranda, M. Caujolle, B. Goncer-Maraver, and M. Sebastian-Viana, ``Estimating the Active and Reactive Power Flexibility Area at the TSO-DSO Interface," \textit{IEEE Transactions of Power Systems,} vol. 33, no. 5, pp. 4741-4750, Sep. 2018.

\bibitem{Florin} F. Capitanescu, ``TSO–DSO Interaction: Active Distribution Network Power Chart for TSO Ancillary Services Provision," \textit{Electric Power Systems Research,} vol. 163, part A, pp. 226-230, 2018.







\bibitem{AC-DSO1}
S. S. Torbaghan, G. Suryanarayana, H. Höschle, R. D'hulst, F. Geth, C. Caerts, and D. Van Hertem, ``Optimal Flexibility Dispatch Problem Using Second-Order Cone Relaxation of AC Power Flows,''  \textit{IEEE Transactions on Power Systems}, vol. 35, No. 1, pp. 98-108, 2020. 

\bibitem{AC-DSO2}
M. Farivar and S. H. Low, ``Branch Flow Model: Relaxations and Convexification - Part I,'' \textit{IEEE Transactions on Power Systems}, vol. 28, no. 3, pp. 2554-2564, Aug. 2013.

\bibitem{Polymeneas} E. Polymeneas and S. Meliopoulos, ``Aggregate Modeling of Distribution Systems for Multi-Period OPF," \textit{2016 Power Systems Computation Conference (PSCC)}, Genoa, 2016, pp. 1-8, doi: 10.1109/PSCC.2016.7540987.

\bibitem{TD4}
Z. Li, Q. Guo, H. Sun, and J. Wang, ``Coordinated Transmission and Distribution AC Optimal Power Flow,'' \textit{IEEE Transactions on Smart Grid}, vol. 9, no. 2, pp. 1228-1240, Jun 2016.

\bibitem{Yuan2018} Z. Yuan, M.R. Hesamzadeh, and D.R. Biggar, ``Distribution Locational Marginal Pricing by Convexified ACOPF and Hierarchical Dispatch," \textit{IEEE Transactions on Smart Grid,} vol. 9, no. 4, pp. 3133–3142, 2018.

\bibitem{SaintPierre2017} A. Saint-Pierre, and P. Mancarella, ``Active Distribution System Management: A Dual Horizon Scheduling Framework for DSO/TSO Interface Under Uncertainty," \textit{IEEE
Transactions on Smart Grid,} vol. 8, no. 5, 2186–2197, 2017.

\bibitem{Rossi} M. Rossi, et al., ``Testing TSO-DSO Interaction Schemes for the Participation of Distributed Energy Resources in the Balancing Market: the Smartnet Simulator," \textit{Presented at the 25th International Conference on Electricity Distribution,} Madrid,
Spain, 2019.

\bibitem{Cadre} H. Le Cadre, I. Mezghani, A. Papavasiliou, ``A Game-Theoretic Analysis of Transmission-Distribution System Operator Coordination," \textit{European Journal of Operational Research}, vol. 274, no. 1, pp. 317–339, 2019.

\bibitem{Mohammadi} A. Mohammadi, M. Mehrtash, A. Kargarian, ``Diagonal Quadratic Approximation for Decentralized Collaborative TSO+DSO Optimal Power Flow," \textit{IEEE Transactions on Smart Grid,} vol. 10, no. 3, pp. 2358–2370, 2019.

\bibitem{Papa2018} A. Papavasiliou, and I. Mezghani, ``Coordination Schemes for the Integration of Transmission and Distribution System Operations," \textit{Presented at the Power Systems
Computation Conference (PSCC)}, Dublin, Ireland, 2018.

\bibitem{TD2}
Z. Li, Q. Guo, H. Sun, and J. Wang, ``Coordinated Economic Dispatch of Coupled Transmission and Distribution Systems Using Heterogeneous Decomposition,'' \textit{IEEE Transactions on Power Systems}, vol. 31, no. 6, pp. 4817-4830, Nov. 2016.

\bibitem{TD3}
Z. Li, Q. Guo, H. Sun, and J. Wang, ``A New LMP-Sensitivity-Based Heterogeneous Decomposition for Transmission and Distribution Coordinated Economic Dispatch''  \textit{IEEE Transactions on Smart Grid}, vol. 9, no. 2, pp. 931-941, May 2016.

\bibitem{Caramanis_2015}
M. Caramanis, E. Ntakou, W. W. Hogan, A. Chakrabortty and J. Schoene, ``Co-Optimization of Power and Reserves in Dynamic T\&D Power Markets With Nondispatchable Renewable Generation and Distributed Energy Resources,''\textit{ Proceedings of the IEEE}, vol. 104, no. 4, pp. 807-836, April 2016.



\bibitem{TSODSOPES18} 
M. A. Bragin, and Y. Dvorkin, ``
Toward Coordinated Transmission and Distribution Operations," 
\textit{In proceedings of the IEEE Power and Energy Society 2018 General Meeting}, Portland, Oregon

\bibitem{Kargarian} A. Kargarian, and Y. Fu, ``System of Systems Based Security-Constrained Unit Commitment Incorporating Active Distribution Grids," \textit{IEEE Transactions of Power Systems}, vol. 29, no. 5, pp. 2489–2498, Sep. 2014.

\bibitem{Nawaz} A. Nawaz, H. Wang, Q. Wu, and M. Kumar Ochani, ``TSO and DSO with Large-Scale Distributed Energy Resources: A Security Constrained Unit Commitment Coordinated Solution," \textit{International
Transactions on Electrical Energy Systems,} vol. 30, no. 3, 2020. https://doi.org/10.1002/2050-7038.12233

\bibitem{Givi} A. Givisiez, K. Petrou, and L. Ochoa, ``A Review on TSO-DSO Coordination Models and Solution Techniques," \textit{Electric Power Systems Research,} vol. 189, 2020 . doi:10.1016/j.epsr.2020.106659


\bibitem{SLR} 
M. A. Bragin, P. B. Luh, J. H. Yan, N. Yu, and G. A. Stern
``Convergence of the Surrogate Lagrangian Relaxation Method," \textit{Journal of Optimization Theory and Applications,} vol. 164, no. 1, 2015, pp. 173-201, DOI: 10.1007/s10957-014-0561-3.

\bibitem{SALR} X. Sun, P. B. Luh, M. A. Bragin, Y. Chen, J. Wan, and F. Wang, ``A Decomposition and Coordination Approach for Large-Scale Security Constrained Unit Commitment Problems with Combined Cycle Units," \textit{IEEE Transactions on Power Systems,} vol. 33, no. 5, pp. 5297-5308, Sept. 2018. DOI: 10.1109/PESGM.2017.8274098

\bibitem{SAVLR} 
M. A. Bragin, P. B. Luh, B. Yan, and X. Sun. 
``A Scalable Solution Methodology for Mixed-Integer Linear Programming Problems Arising in Automation," 
\textit{IEEE Transactions of Automation Science and Engineering}, vol. 16, no. 2, pp. 531 - 541, Jun. 2018. 

\bibitem{Deniz} D. Gurevin, S. Zhou, L. Pepin, B. Li, M. A. Bragin, C. Ding, and F. Miao, ``A Surrogate Lagrangian Relaxation-Based Model Compression for Deep Neural Networks." https://arxiv.org/pdf/2012.10079.pdf

\bibitem{AC_rec_1} G. L. Torres, V. H. Quintana, and G. Lambert-Torres, ``Optimal Power Flow on Rectangular Form Via an Interior Point Method," \textit{in IEEE North American Power Symposium,} 1996.

\bibitem{AC_rec_2} V. da Costa, N. Martins, and J. L. R. Pereira, ``Developments in the Newton Raphson Power Flow Formulation Based on Current Injections," \textit{IEEE Transactions on Power Systems,} vol. 14, no. 4, pp. 1320–1326, November 1999.

\bibitem{AC_rec_3} X. P. Zhang, S. G. Petoussis, and K. R. Godfrey, ``Nonlinear Interior Point Optimal Power Flow Method Based on a Current Mismatch Formulation," \textit{IEEE Proceedings-Generation, Transmission and Distribution,} vol. 152, no. 6, pp. 795–805, November 2005.













\bibitem{RectCoord1}
X. Bai, H.Wei, K. Fujisawa, and Y.Wang, ``Semidefinite Programming
for Optimal Power Flow problems," \textit{International Journal of Electrical Power and Energy Systems}, vol. 30, no. 6-7, pp. 383-392, 2008.

\bibitem{RectCoord2}
Q. Li , L. Yang and S. Lin, ``Coordination Strategy for Decentralized Reactive Power Optimization Based on a Probing Mechanism,`` \textit{IEEE Transactions on Power Systems}, vol. 30, no. 2, pp. 555-562, Mar. 2015. 

\bibitem{matpower} R. D. Zimmerman, C. E. Murillo-Sanchez (2020). MATPOWER (Version 7.1) [Software]. Available: https://matpower.org. Accessed: 01/23/21

\bibitem{Marten} \textcolor{blue}{F. Marten, L. Lwer, J. C. Tbermann, M, Braun, ``Optimizing the reactive power balance between a distribution and transmission grid through iteratively updated grid equivalents,'' In \textit{Power systems computation conference (PSCC),}, pp. 1–7, 2014.}

\bibitem{Nik} \textcolor{blue}{A. Nikpour, A. Nateghi, M. Shafie-khah, and J. P. Catalão, ``Hybrid stochastic risk-based approach for a microgrid participating in coupled active and reactive power market,'' \textit{International Journal of Electrical Power \& Energy Systems}, 131, 107080.}

\bibitem{Gil} \textcolor{blue}{J. B. Gil, T. G. San Román, J. A. Rios, and P. S. Martin, ``Reactive power pricing: a conceptual framework for remuneration and charging procedures,'' \textit{IEEE Transactions on Power Systems}, vol 15, no. 2, pp. 483-489, 2000.}

\bibitem{Lipka} \textcolor{blue}{P. Lipka, S. S. Oren, R. P. O’Neill, and A. Castillo, ``Running a more complete market with the SLP-IV-ACOPF,'' \textit{IEEE Transactions on Power Systems,} vol. 32, no. 2, pp. 1139-1148, 2016.}








\bibitem{BingRamp}
B. Yan, P. B. Luh, E. Litvinov, T. Zheng, D. Schiro, M. A. Bragin, F. Zhao, J. Zhao, and I. Lelic, ``A Systematic Formulation Tightening Approach for Unit Commitment Problems," \textit{IEEE Transactions on Power Systems,} vol. 35, no. 1, pp. 782-794, Aug. 2019. 





\bibitem{ABB}
N. Raghunathan, M. A. Bragin, B. Yan, P. B. Luh, K. Moslehi, X. Feng, Y. Yu, C.-N. Yu, and C.-C. Tsai, ``Exploiting Soft Constraints within Decomposition and Coordination Methods for Sub-Hourly Unit Commitment,” \textit{TechRxiv. Preprint. https://doi.org/10.36227/techrxiv.12950414.v1}




\end{thebibliography}
\end{document}